\documentclass[aps,prb,nofootinbib,tightenlines,twocolumn,floatfix,superscriptaddress,showpacs,longbibliography]{revtex4-1}
\usepackage{graphicx}
\usepackage{amsmath}
\usepackage{amssymb}
\usepackage{hyperref}
\usepackage{soul}
\usepackage[pdftex]{color}

\DeclareMathOperator{\sech}{sech}

\newcommand{\ket}[1]{|#1 \rangle}

\def\Tr{\mathrm{Tr}} 

\newcommand{\be}{\begin{eqnarray}}
\newcommand{\ee}{\end{eqnarray}}
\newcommand{\bbm}{\begin{bmatrix}}
\newcommand{\ebm}{\end{bmatrix}}
\newcommand{\bpm}{\begin{pmatrix}}
\newcommand{\epm}{\end{pmatrix}}
\renewcommand{\v}[1]{{\bf #1}}
\newcommand{\nn}{\nonumber \\}

\begin{document}

\title{Perfect transmission and Aharanov-Bohm oscillations in topological insulator nanowires with nonuniform cross section}
\author{Emmanouil Xypakis}
\affiliation{Max-Planck-Institut f{\"u}r Physik komplexer Systeme, 01187 Dresden, Germany}
\author{Jun-Won \surname{Rhim}}
\affiliation{Max-Planck-Institut f{\"u}r Physik komplexer Systeme, 01187 Dresden, Germany}
\affiliation{Center for Correlated Electron Systems, Institute for Basic Science (IBS), Seoul 08826, Korea}
\affiliation{Department of Physics and Astronomy, Seoul National University, Seoul 08826, Korea}
\author{Jens H. \surname{Bardarson}}
\affiliation{Max-Planck-Institut f{\"u}r Physik komplexer Systeme, 01187 Dresden, Germany}
\affiliation{Department of Physics, KTH Royal Institute of Technology, Stockholm, SE-106 91 Sweden}
\author{Roni Ilan}
\affiliation{Raymond and Beverly Sackler School of Physics and Astronomy, Tel-Aviv University, Tel-Aviv 69978, Israel}
\begin{abstract}

Topological insulator nanowires with uniform cross section, combined with a magnetic flux, can host both a perfectly transmitted mode and Majorana zero modes.
Here we consider nanowires with rippled surfaces---specifically, wires with a circular cross section with a radius varying along its axis---and calculate their transport properties.
At zero doping, chiral symmetry places the clean wires (no impurities) in the AIII symmetry class, which results in a $\mathbb{Z}$ topological classification. 
A magnetic flux threading the wire tunes between the topologically distinct insulating phases, with perfect transmission obtained at the phase transition. 
We derive an analytical expression for the exact flux value at the transition.
Both doping and  disorder breaks the chiral symmetry and the perfect transmission.
At finite doping, the interplay of surface ripples and disorder with the magnetic flux modifies quantum interference such that the amplitude of Aharonov-Bohm oscillations reduces with increasing flux, in contrast to wires with uniform surfaces where it is flux-independent.
\end{abstract}
\maketitle
\section{Introduction}

Three dimensional topological insulators (TIs) are bulk insulators supporting a topologically protected metallic Dirac-like surface state~\cite{Moore2010,Hasan2010,Qi2011}.
Spin-momentum locking strongly affects the transport properties of these materials, since closed electron trajectories contribute to quantum interference with a $\pi$ Berry phase~\cite{Bardarson2013}.
In particular, when such a system is realized in the confined geometry of a nanowire, electrons accumulate the $\pi$ Berry phase when looping the wire, resulting in a gapped surface-state spectrum~\cite{Lee2009,Zhang2010,Bardarson2010}.
A magnetic flux of $\phi_0 = h/e$ through a uniform nanowire exactly compensates this phase and eliminates the surface-state gap~\cite{Ran2008,Rosenberg2010,Bardarson2010,Zhang2010}.
The closing of the gap is accompanied by the appearance of a perfectly transmitted mode which plays a central role in the physics of TI nanowires~\cite{Bardarson2013}.

The parity of the number of modes is reflected in transport.
Theoretically, the normal state conductance oscillates with the magnetic flux with a period of $h/e$ for weak disorder and $h/2e$ for strong disorder, with a magnetic-field-independent amplitude~\cite{Bardarson2010}.
Experiments observe such magneto-oscillations with a period consistent with the theoretical predictions, but with an amplitude that decays with increasing magnetic field strength~\cite{Zhang2009,Chen2015,Dufouleur2017}.
Moreover, the odd number of modes is a crucial feature underlying the emergence of topological superconductivity in wires coupled to superconductors via the proximity effect~\cite{Cook2011}.
At a normal-superconducting interface, the perfectly transmitted mode can be split into Majorana modes realizing a Majorana interferometer~\cite{DeJuan2014}. 

The perfectly transmitted mode and its consequences, while robust to scalar disorder, rely on the uniformity of the wire in order to have a constant flux through it.
This raises the question of what effects ripples in the surface, leading to a cross section that varies along the wire, have on transport properties in TI nanowires.
Such curvature effects are well studied in various systems:
in graphene, for example, smooth variations on microscopic length scales (such as surface ripples, corrugations or strain) introduce local curvature into the low energy theory of Dirac fermions via fictitious gauge fields~\cite{DeJuan2011,Levy2010}. 
These gauge fields lead to huge pseudomagnetic fields that can drastically alter the transport properties~\cite{Vozmediano2010,Amorim2016,Jeong2011,Yan2013,Zwierzycki2014,Burgos2015,Guan2017}.
Effective gauge fields are similarly induced by strain in topological crystalline insulators~\cite{Tang2014a} and Weyl semimetals~\cite{Cortijo2015}, but such effects are less studied in TI nanowires~\cite{Zhang2010,Takane}.
Since a change in the circumference of a wire leads to a local variation in the magnetic flux, the emergent time-reversal symmetry at half a flux quantum is absent. 
The varying flux therefore has the potential to localize the perfectly transmitted mode and eliminate the chance for robust transport or the formation of Majorana fermions.

\begin{figure}[tb!]
    \centering
    \includegraphics[width=0.9\columnwidth]{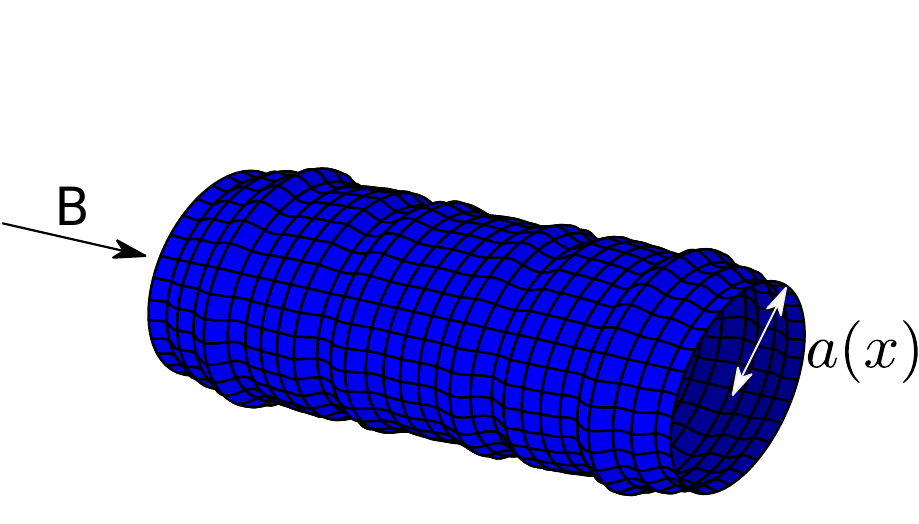}
    \caption{A schematic plot of a topological insulator nanowire with a rippled surface. The radius of the wire $a(x)$ fluctuates in the longitudinal direction around a mean radius $a_0$ with a disk cross section at every $x$. A magnetic field $B$ is applied along the symmetry axis.
    }
    \label{fig:schema}
\end{figure}

In this work we address the above-mentioned questions, by studying how random ripples in the surface of TI nanowires (see Fig.~\ref{fig:schema}) affect its transport properties.
The curved wire geometry is modeled by an intrinsic gauge field in the Dirac equation, which originates from the spin connection~\cite{Utiyama1956,Takane}.
This gauge field preserves the chiral symmetry of the clean wire at the Dirac point---in contrast to scalar disorder which breaks it---and places the wire in the AIII symmetry class characterized by a $\mathbb{Z}$ topological index~\cite{Schnyder2009a,Chiu2016}. 
At the Dirac point for a clean wire we can change the topological index to any $\mathbb{Z}$ value by tuning the magnetic field, and obtain a perfect transmission at the phase transition between topologically distinct classes.
Both scalar disorder and a finite chemical potential break the chiral symmetry and lead to a reduction in the conductance.
Similarly, the combination of ripples and scalar disorder randomizes the interference between distinct paths winding the wire, which results in a decreasing amplitude of magneto-oscillations with increased magnetic flux. 

\section{Model}
We assume the bulk of the topological insulator wire to be a perfect insulator, consistent, for example, with recent experiments on BiSbTeSe$_2$ nanoribbons~\cite{Jauregui2018,Kayyalha:2019gk}. Even in materials with conducting bulk, the surface state contribution can be isolated and is unaffected by the bulk~\cite{Cai:2018dla}.
We model the metallic boundary of a wire of length $L$ as a rotationally symmetric two-dimensional compact manifold with radius $a(x)$ that varies along the length of the wire. 
This surface can be described with the $2+1$D metric (in natural units)
\begin{equation}
    g_{\mu\nu} (x) = \textrm{diag}[-1,\, 1+ (\partial_x a)^2, \, a^2],
	\label{radius_metric0}
\end{equation}
where the index $\mu,\nu=t,x,\theta$ runs over the temporal coordinate $t$ and the spatial coordinates $x$ running along the length of the wire and $\theta$ the angle of rotation about the symmetry axis.
The nanowire is further coupled to a coaxial magnetic field $\mathbf{B}$ as shown in Fig.~\ref{fig:schema}.

Assuming the nanowire's bulk to be a perfect insulator~\cite{Xypakis2017,Hasan2010,Moore2010,Qi2011,Bardarson2013}, 
the effective Hamiltonian for the geometry \eqref{radius_metric0} takes the form (see App.~\ref{sec:DiracEquation} for derivation)
\begin{eqnarray}
    H &=& v_F [\sigma_x g_{xx}^{-1/2} (-i\hbar\partial_x + A_x) \nn
    &+ & \sigma_yg_{\theta\theta}^{-1/2} (-i \hbar\partial_{\theta} + eA_\theta) ].
	\label{eq:Hamiltonian}
\end{eqnarray}
Here $v_F$ is the Fermi velocity, approximately equal to $5 \times 10^5$~m/s in Bi$_2$Te$_3$ and Bi$_2$Se$_3$~\cite{Chen2015,Zhang2009}, 
and $\sigma_x$ and $\sigma_y$ are the Pauli matrices.
The two vector potentials
\begin{eqnarray}
	A_{\theta} = \frac{Ba^2(x)}{2}, \quad
	A_x = \frac{-i\hbar \partial_x a(x)}{2a(x)},
	\label{vector potential}
\end{eqnarray}
should not be thought of as the components of a single vector potential, even though we choose to denote them as if they were. 
Instead, their physical origin is distinct:  
$A_{\theta}$ describes the minimal coupling of the Dirac equation with the (three-dimensional) magnetic field $\boldsymbol{B}$,
while $A_x$ is a spin connection that takes the form of an imaginary curvature-induced gauge potential, and ensures that the Hamiltonian is Hermitian
\begin{equation}
    \int_{M} dx d\theta \sqrt{g} \Psi^{\dagger} H \Psi =\int_{M} dx d\theta \sqrt{g} (H\Psi)^{\dagger}  \Psi,
    \label{Hermiticity}
\end{equation}
where $g$ is the determinant of the metric.
The $2\pi$ rotation of the momentum-locked spinor around the symmetry axis is incorporated in the antiperiodic boundary condition
$\Psi (x,\theta + 2\pi)= -\Psi(x,\theta)$ \cite{Rosenberg2010,Bardarson2010,Zhang2010}.
The fluctuations  $\delta a(x)= a(x)- a_0$ around the mean radius $a_0$ are taken to be Gaussian correlated 
\begin{equation}
	\langle \delta a(x) \delta a(x') \rangle = 
	\delta_a ^2 \exp\left(\frac{-|x -x '|^2}{\xi_a^2}\right),
	\label{disorder_a}
\end{equation}
where $\langle \dots \rangle$ denotes the average, $\delta_a$ is the amplitude of radius fluctuations, and $\xi_a$ is the characteristic ripple length; in this work we take $a_0 =30$nm.
The scalar disorder potential $V(x, \theta)$ is also taken to be Gaussian correlated
\begin{equation}
	\langle V(x,\theta)V(x',\theta') \rangle = 
    K_V\frac{ (\hbar v_F)^2 }{2\xi_V^2}e^{\frac{-(x-x')^2 -a_0^2 (\theta-\theta')^2}{2\xi^2 _V}},
	\label{disorderV}
\end{equation}
with amplitude $K_V$ and correlation length $\xi_V$.
In all the data in this paper we take the wire length $L = 300\mathrm{nm} \gg \xi_V,\xi_a$.

We numerically calculate the conductance for a given surface geometry and disorder realization via the Landauer formula $G=(e^{2}/h) \Tr \;t^{\dagger} t$, with $t$ the matrix of transmission amplitudes, using the procedure described, e.g., in Refs.~\cite{Beenakker1997, Mello1988,Tworzydo2006, Xypakis2017}.
We then average over an ensemble of $1000$ different realizations.
A subtlety in the calculation is in the normalization of the transverse modes, which arises since the radius of the nanowire depends on the $x$ coordinate.
From the solutions $\Psi$ of the Schr{\"o}dinger equation $H\Psi = E_F\Psi$ with $H$ the Hamiltonian~\eqref{eq:Hamiltonian} and $E_F$ the Fermi energy, the transport modes $\varphi$ carrying unit current are obtained (see App.~\ref{sec:Transfer} for derivation) as
\begin{equation}
    \varphi_n(x)  = \sqrt{2\pi a (x)} R \int_{0}^{2 \pi} d\theta e^{-i(n-\frac{1}{2}) \theta}  \Psi(x, \theta) ,
    \label{transportmode}
\end{equation}
where the rotation matrix $R = (\sigma^x + \sigma^z)/\sqrt{2}$, and the current operator $\hat{j}_x = \partial H/\partial p_s = \sigma^x$ with $p_s = [1+(\partial_x a)^2]^{-1/2}\partial_x$ the momentum parallel to the nanowire surface along the length of the wire. 
In this basis, the transfer matrix $T_{m,n}$, defined from $\varphi_m (L) = T_{m,n} \varphi_n (0)$, is found to be
\begin{eqnarray}
    T_{m,n} & = & \mathcal{P}_{x_s}  \exp\left[ \int_{0}^{L}dx_s\  M_{m,n}(x_s) \right], \nonumber\\
    M_{m,n}(x) & = & \frac{i}{\hbar v_F}[ E_F \delta_{n,m} - V_{nm}(x) ] \sqrt{1+(\partial_x a)^2} \sigma^z\nonumber \\
    & + & \delta_{n,m}\frac{\sqrt{1+(\partial_x a)^2}}{a}(n -\frac{1}{2} +e\frac{B a^2}{2\hbar})\sigma^x ,
    \label{transfermatrix}
\end{eqnarray}
where $\mathcal{P}_{x_s}$ is the path ordering operator and $V_{nm}= \frac{1}{2 \pi} \int_{0}^{2\pi} d\theta e^{i(n-m)\theta}V(x, \theta)$.
From the transfer matrix one obtains the transmission matrix $t$ and the conductance.

\section{Chiral symmetry at zero doping}
A clean but rippled nanowire at zero doping is chiral symmetric, $\sigma^z H \sigma^z = -H$, with respect to the Dirac point, and therefore belongs to the symmetry class AIII~\cite{Schnyder2009a,Chiu2016}.
AIII has a $\mathbb{Z}$ topological classification and in the nanowires we can tune between the distinct topological phases with the magnetic flux $\phi$.
A perfectly transmitted mode accompanies the transition between distinct phases at the Dirac point, and gives rise to a quantized conductance at the critical points.

\begin{figure}[tb]
	\begin{center}
		\includegraphics[width=\columnwidth]{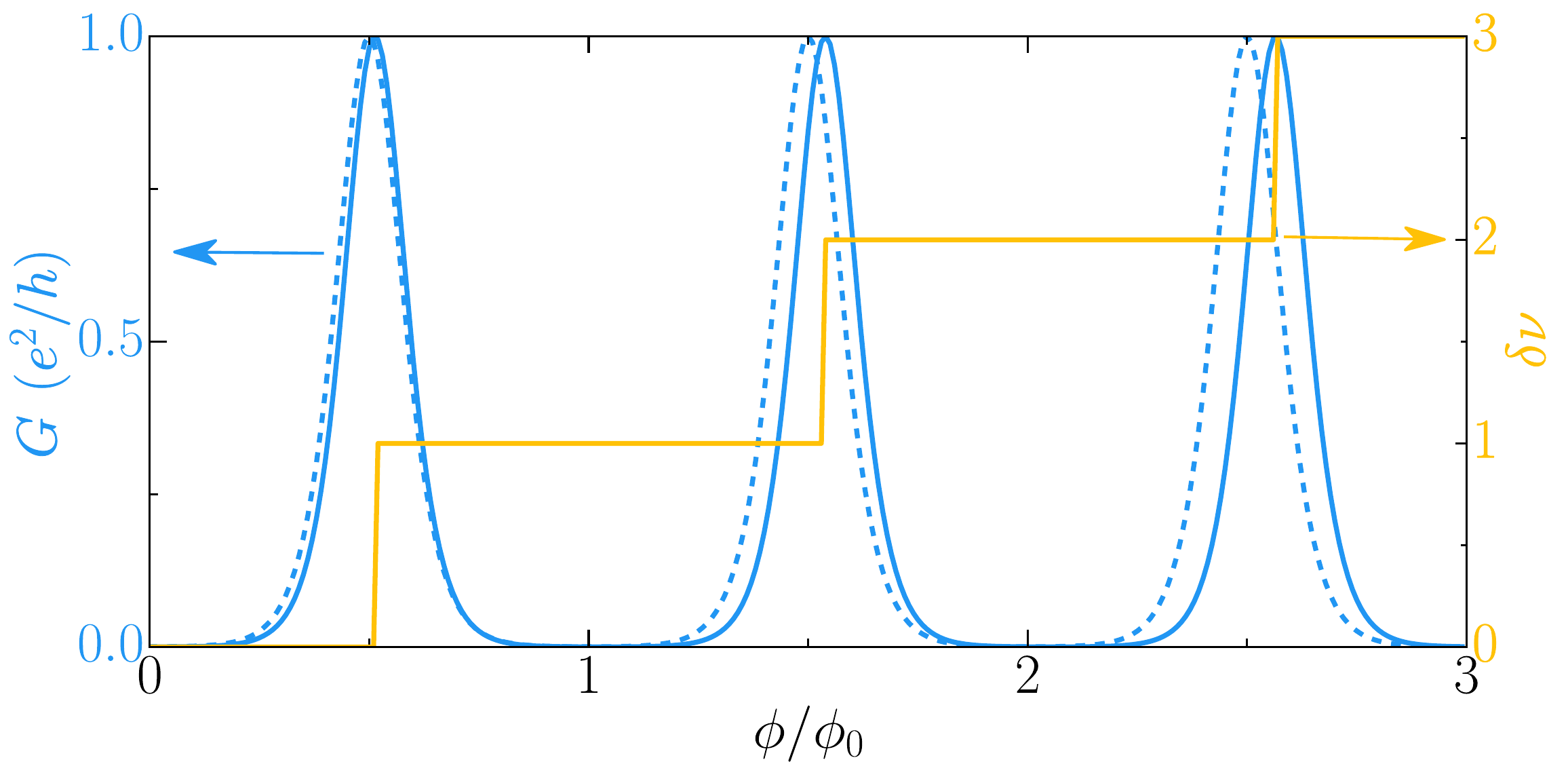}
	\end{center}
    \caption{The conductance $G$ (left axis) and the topological index $\delta \nu = \nu (\phi) - \nu (\phi = 0)$ (right axis) for a single surface realization of a clean TI nanowire at the Dirac point $E_F = 0$ and at $\delta_a = 5 $nm (solid line), as a function of the magnetic flux  $\phi / \phi_0 = Ba_0^2$. The conductance peaks at $G=e^2/h$ at the fluxes $\phi_{\textrm{max}, n}$ given by the Eq. \eqref{flux_max}. The dotted line shows for comparison the conductance of a uniform wire with $\delta_a = 0$.}
	\label{fig:figure2}
\end{figure}
To demonstrate this, we plot in Fig.~\ref{fig:figure2} the conductance $G$ at the Dirac point of a clean rippled nanowire,
as a function of the magnetic flux $\phi/ \phi_0 = B a_0^2 /2$.
The conductance peaks at one conductance quantum at magnetic flux values given by 
\begin{equation}
    \frac{\phi_{\textrm{max},n}}{\phi_0} = a_0^2\left(n+ \frac{1}{2}\right)\frac{  \int_{0}^{L} dx~ a^{-1} \sqrt{1 + (\partial_x a)^2}}{\int_{0}^{L} dx~ a \sqrt{1 + (\partial_x a)^2}},
	\label{flux_max}
\end{equation} 
which are obtained from the transfer matrix \eqref{transfermatrix} and depends on the specific nanowire geometry (see App.~\ref{sec:Transfer}).
This expression reduces to $\phi_{\textrm{max},n}/\phi_0 = n+1/2$ for uniform wires.
When the nanowire radius fluctuates smoothly, $\partial_x a \ll 1$ or on average $\delta \ll \xi $, 
the formula above is simplified and the maximum is shifted compared to the uniform cylinder by 
$\delta\phi/\phi_0 = (n+1/2)\int_{0}^{L} dx (\delta a(x) / a_0)^2/L$.
By averaging over surface realizations, this shift takes the form
\begin{equation}
    \frac{\langle \delta\phi\rangle}{\phi_0} =  \frac{\delta_a^2}{a_0^2} \left(n+\frac{1}{2}\right),
    \label{phimax2}
\end{equation}
consistent with numerical results (data not shown).
The perfectly transmitted mode indicates a topological phase transition.
For the AIII symmetry class in one dimension the topological index $\nu$ that characterizes the system is equal to the number of negative eigenvalues of the reflection matrix $r$ \cite{Fulga2011a,Fulga2011,Altland1997}.
In Fig.~\ref{fig:figure2}, we plot this topological index $\nu$ for the same parameters.
The conductance maxima at $\phi_\textrm{max,n}$ occur exactly at the points where $\nu$ jumps by one; since at these points one of the reflection eigenvalues changes sign, it must be equal to zero at the transition, resulting in a perfectly transmitted mode. We note that even if for any given disorder realization a perfect conductance is obtained at some value of the flux, if the conductance is averaged over, one would expect it to decay as a function of the system size. This is discussed in more detail in App.~\ref{sec:DisorderAverage}.

\begin{figure}[tb]
	\begin{center}
		\includegraphics[width=\columnwidth]{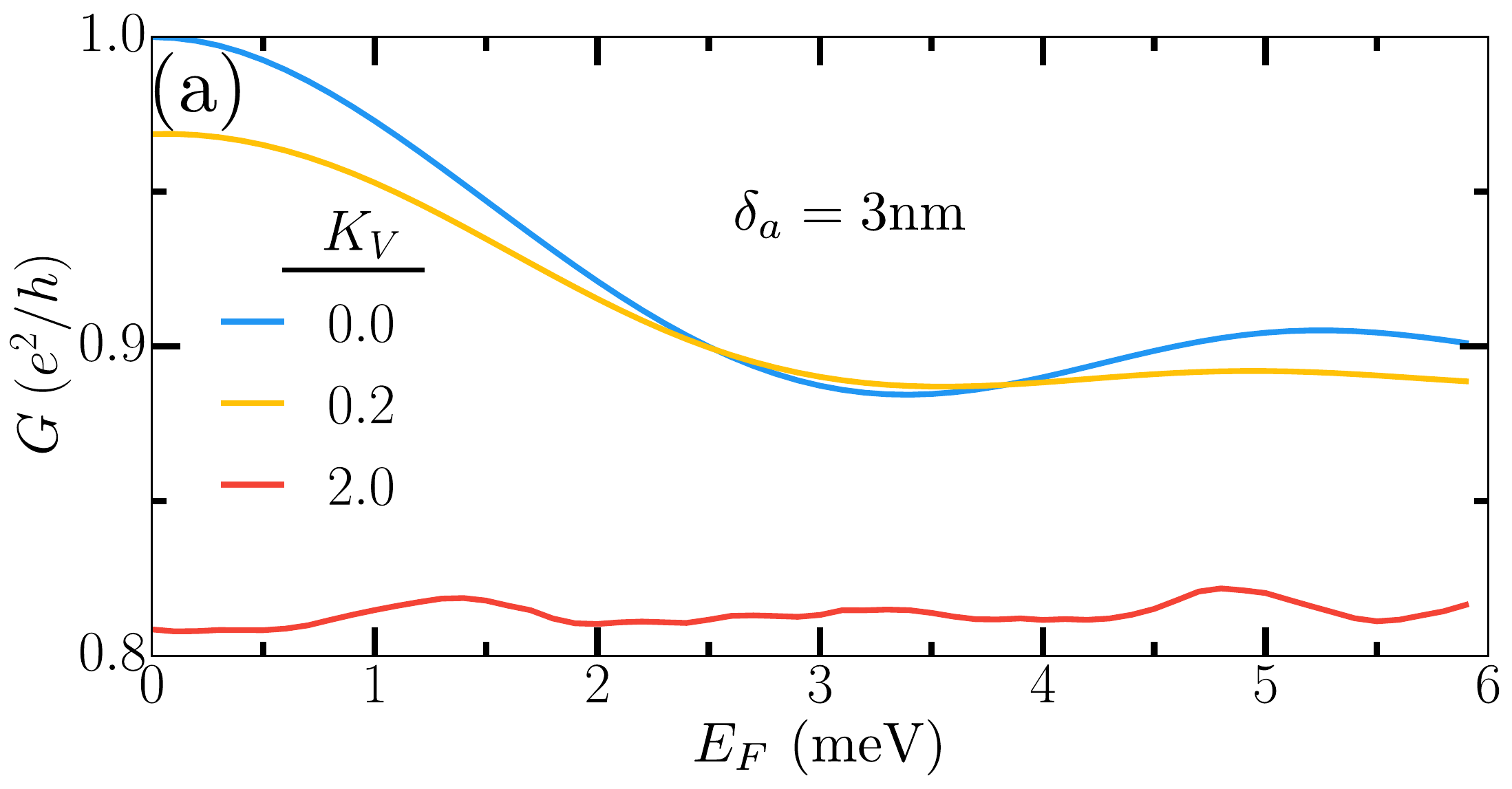}
		\includegraphics[width=\columnwidth]{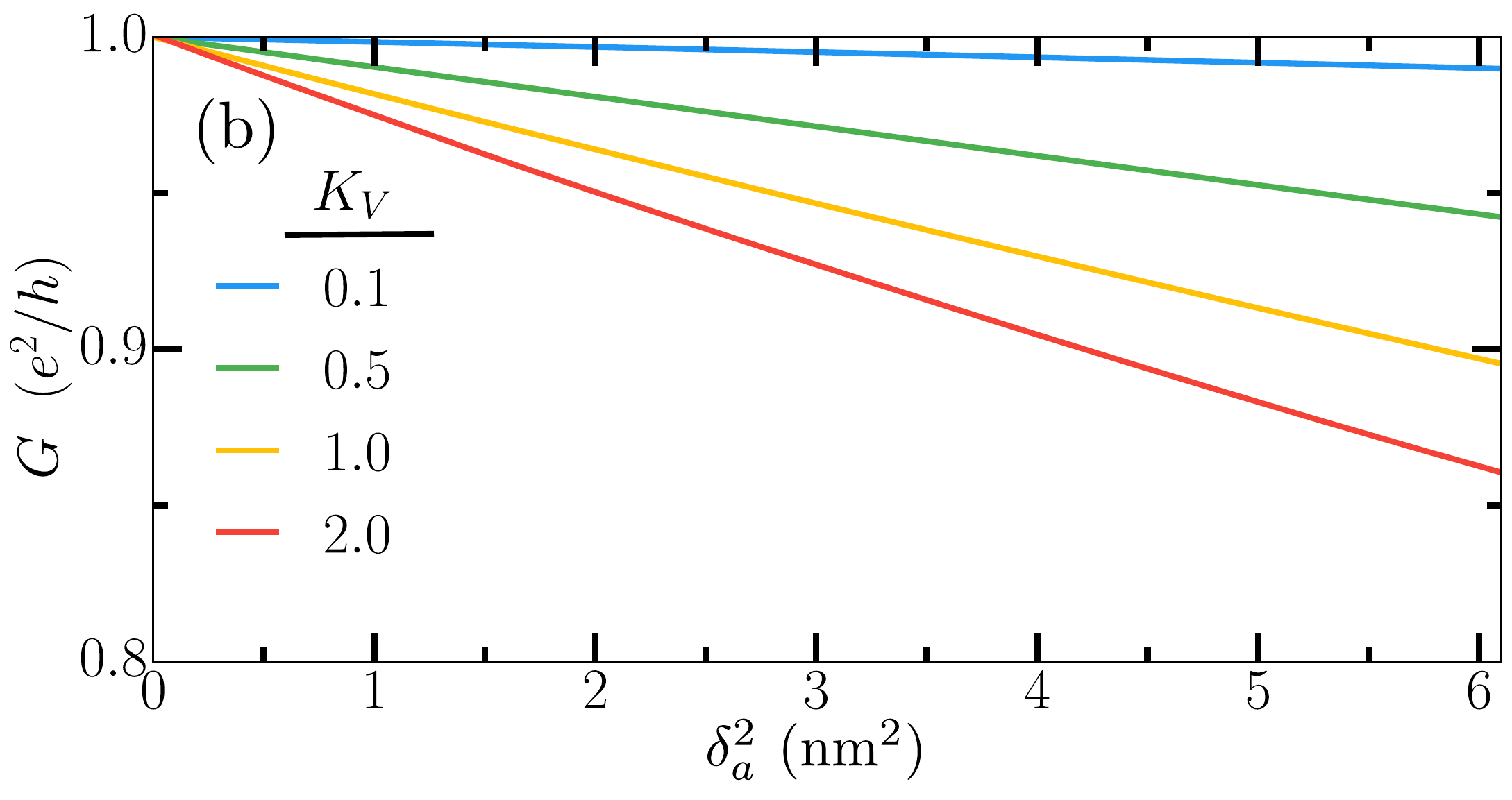}
		\includegraphics[width=\columnwidth]{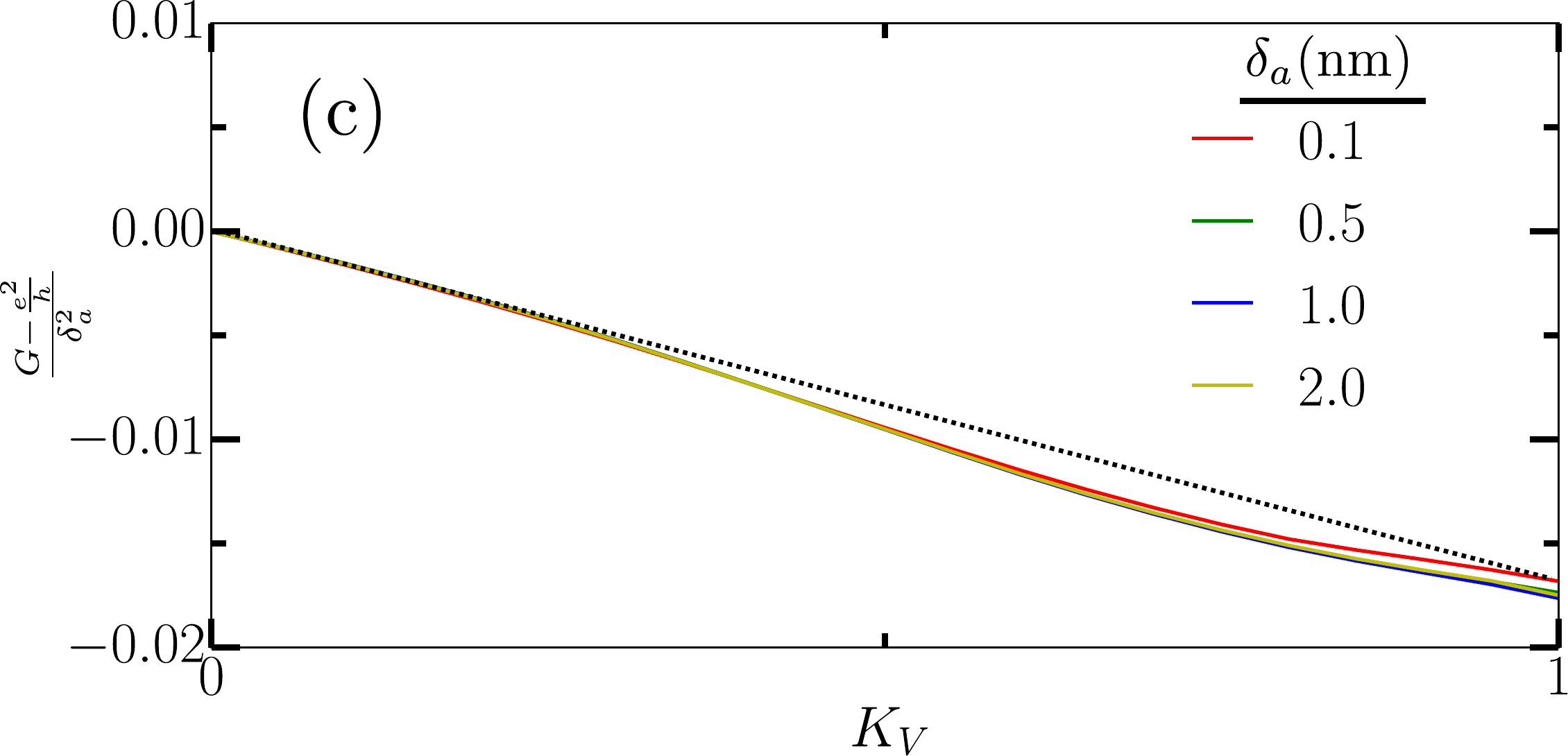}
	\end{center}
    \caption{Conductance $G$, averaged over disorder and geometry at a fixed flux $\phi/ \phi_0= 1/2 + \delta_a^2/2a_0^2$, as a function of (a) chemical potential $E_F$;
        (b) ripple amplitude $\delta_a^2$ at the Dirac point $E_F=0$; and (c) disorder strength $K_V$ at the Dirac point $E_F=0$. (a) and (b) are shown for different disorder strengths $K_V$ while (c) is shown for different ripple amplitudes $\delta_a$; the dashed line is a linear fit to the numerical data at small $K_V$.
	}
	\label{fig:figure3}
\end{figure}
\begin{figure}[tb]
	\begin{center}
		\includegraphics[width=\columnwidth]{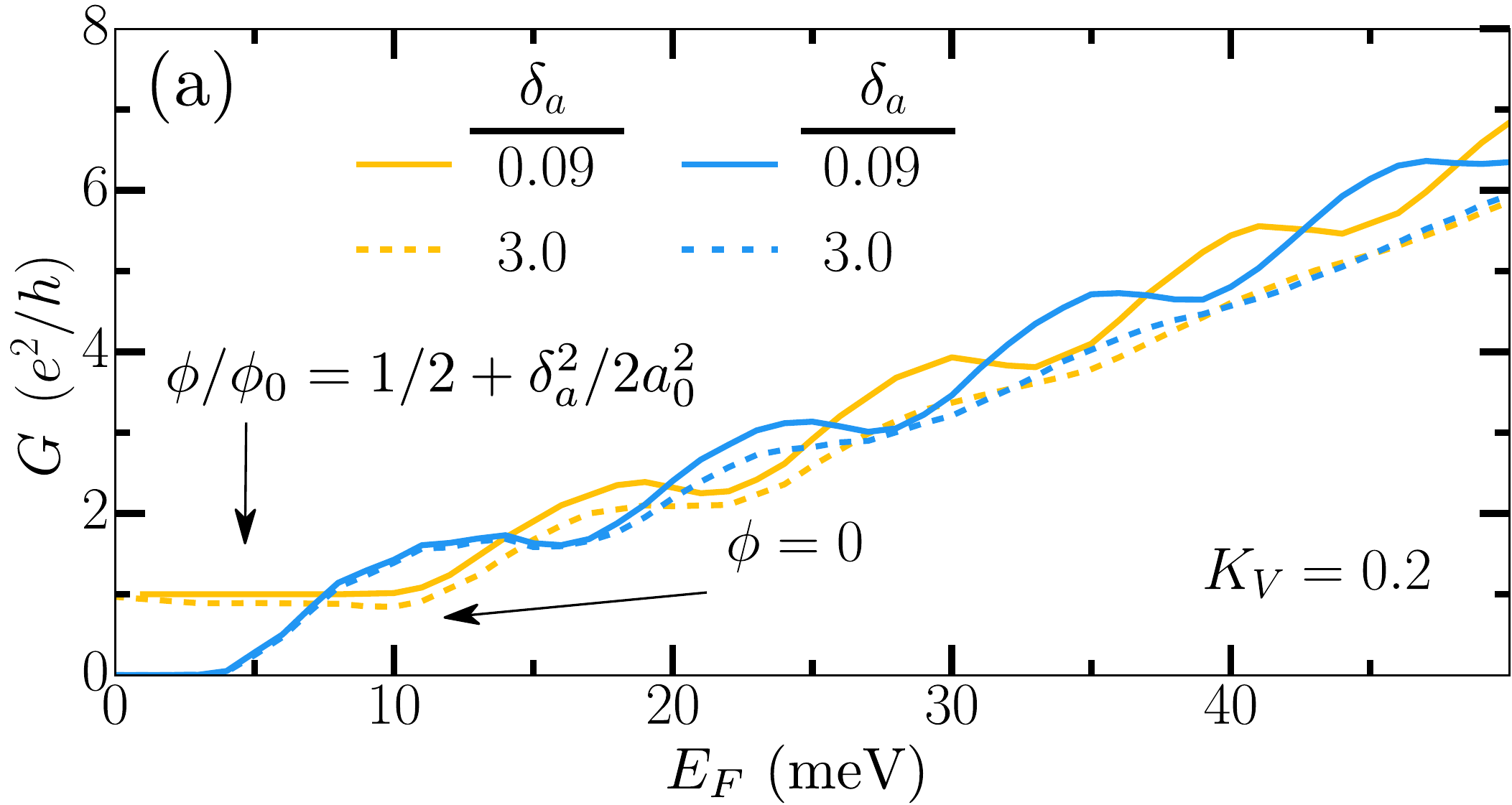}
		\includegraphics[width=\columnwidth]{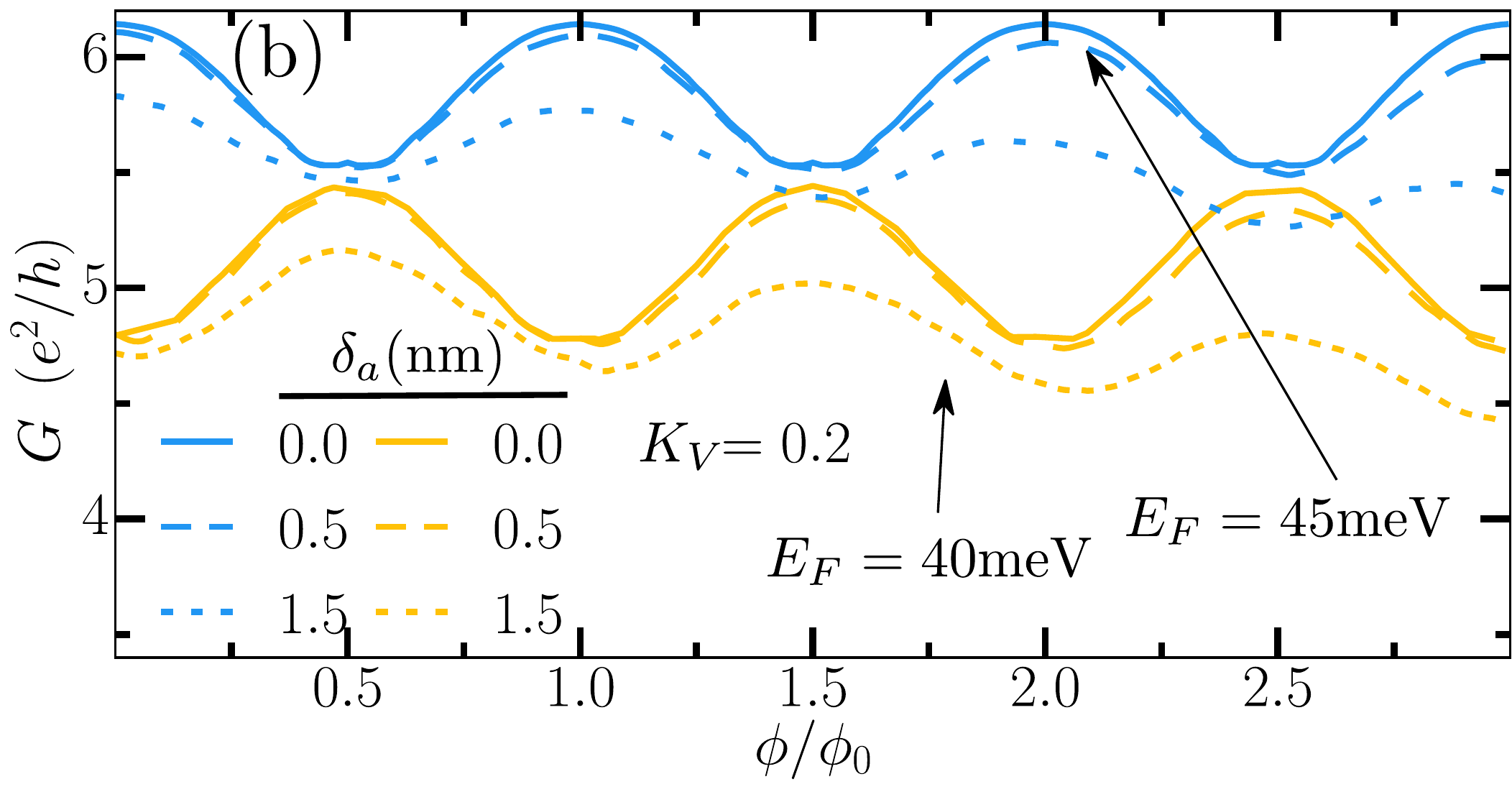}
		\includegraphics[width=\columnwidth]{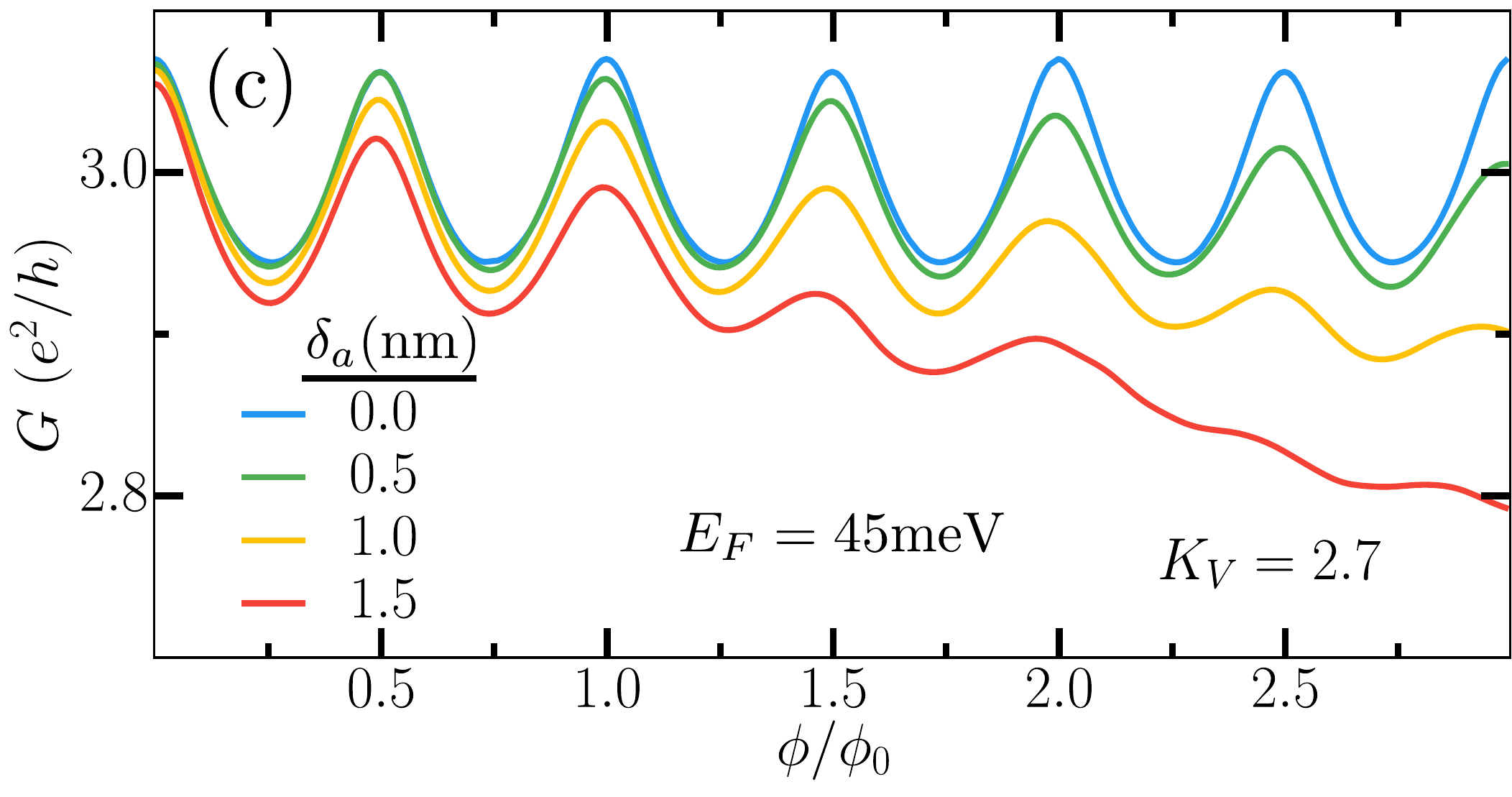}
	\end{center}
	\caption{Average conductance $G$ as a function of (a) Fermi energy $E_F$, for two different values of flux and different ripple amplitudes; and as a function of flux at (b) two values of the Fermi energy far from the Dirac point and at weak disorder $K_V=0.2$, and (c) one value of the Fermi energy and large disorder $K_V=2.7$. (b) and (c) are shown for different ripple amplitudes.
}
	\label{fig:figure4}
\end{figure}

\section{Chiral symmetry breaking at finite doping}
Both scalar disorder potential $V$ and finite doping $E_F \neq 0$ break the chiral symmetry (Fig.~\ref{fig:figure3}).
In contrast to the uniform wire ($ \delta_a = 0$), where there is a perfectly transmitted mode at finite energies---due to the effective time-reversal symmetry obtained at half a flux quantum---in the rippled wires we obtain such a mode only at the Dirac point.
For a rippled wire,  for three different disorder strengths, the average conductance $G$ drops with increasing doping [Fig.~\ref{fig:figure3}(a)].
For clean rippled wires, the conductance indeed drops below $e^2/h$ at finite energies, while for finite disorder strengths $K_V > 0$ the conductance goes below $e^2/h$ also at the Dirac point.
The reduction of the conductance is proportional to the disorder strength.

In Fig.~\ref{fig:figure3}(b) we plot the average conductance $G$ at the Dirac point as a function of the ripple amplitude $\delta_a^2$  for different values of $K_V$ and at the fixed flux $\phi=(1/2 + \delta_a^2/2a_0^2)\phi_0 $ given by Eq.~\eqref{phimax2}.
By increasing $\delta_a$ the conductance decreases linearly for different values of the scalar disorder strength, with a slope that increases with increasing $K_V$.
In Fig.~\ref{fig:figure3}(c) we show how the average conductance depends on the scalar disorder $K_V$ for different values of the amplitude of the radius fluctuation $\delta_a$ and at flux $\phi=(1/2 + \delta_a^2/2a_0^2)\phi_0$.
The drop is nonlinear for large $K_V$ and at small $K_V$ it is proportional to $K_V$.
%
Hence, the conductance drop at the phase transition point indicates that the chiral symmetry breaking effect becomes stronger by increasing the amplitude of the radius fluctuations in a disorder wire.

\section{Quantum interference at finite doping}
In uniform wires ($\delta_a = 0$), the conductance at weak disorder shows Aharonov-Bohm oscillations with period $h/e$, with a chemical potential dependent $\pi$-phase shift~\cite{Bardarson2010}; here we discuss the faith of these oscillations in rippled wires.
In Fig.~\ref{fig:figure4}(a) we plot the average conductance $G$ as a function of the chemical potential $E_F$ for two different values of the magnetic flux and two different $\delta_a$ in the weak disorder regime.
For small $\delta_a$, the conductance $G$ oscillates as a function of doping with a period equal to the level spacing $\Delta = h v_F /a_0$.
The local maxima of $\phi/\phi_0=1/2 + \delta^2_a/2a^2_0$ are at the same energies as the local minima of $\phi=0$ and vice versa.
This interference pattern is very similar to that obtained in uniform cylinder nanowires and shows that time-reversal symmetry breaking effect is weak for small $\delta_a$.
In contrast, by increasing $\delta_a$ to one tenth of the radius $a_0$, the amplitude of the oscillations becomes very small at $30$ meV.
That  means that the interference is incoherent, since after a complete loop around the wire the phase accumulated is random.

In Fig.~\ref{fig:figure4}(b), we plot the average conductance $G$  in the ballistic limit as a function of the magnetic flux $\phi$ for two different values of the chemical potential and three different values of $\delta_a$.
For a uniform cylinder we have a perfect $\phi_0$ Aharonov-Bohm magneto-oscillation period in the conductance.
By increasing $\delta_a$, the amplitude of the oscillations decreases at higher magnetic fluxes.
As a result, the Aharonov-Bohm interference period is gradually lost at higher fluxes and the conductance drops.
In Fig.~\ref{fig:figure4}~(c) we study the magneto-oscillation period of the conductance in the diffusive limit, where $K_V$ is large, for different values of $\delta_a$.
In the uniform cylinder $\delta_a=0$ we have a $\phi_0/2$ period.
The maxima of the conductance are understood by weak anti-localization, since, for a perfect wire, at $\phi=n/2 \phi_0$ the system is time reversal symmetric.
%
At $\delta_a \neq 0$ we break time-reversal symmetry universally, except at zero flux.
As a result the height of the  weak anti-localization peaks decrease with increasing $\delta_a$.
Eventually, at large values of $\delta_a$ and $\phi$, the local deviations in magnetic flux due to ripples become large enough that the phase between time reversed paths is randomized, resulting in a loss of quantum interference and the conductance decreases almost monotonically (see also \cite{Sacksteder2016})
Both in the diffusive and the ballistic limit the reduction of the magneto-oscillation amplitude of the conductance is in contrast with the uniform cylinder nanowire, where the amplitude of the oscillations remains the same for all values of the magnetic flux, theoretically up to infinite flux.
The effect is stronger in the diffusive limit when the $\delta_a/a_0$ ratio is small, where essentially any magneto-oscillation signal can only be observed up to two quantum fluxes.

\section{Fluctuations in the azimuthal direction}
Up to this point, we ignored, for simplification, radius fluctuations in the azimuthal $\theta$ direction. In the more general case when the radius fluctuates in both directions, $a=a(x,\theta)$, the Hamiltonian takes the form (see App.~\ref{sec:DiracEquation} for derivation) 
\begin{align}
    H_{\Xi} &= \frac{1}{ \sqrt{1+a_x^2}} \sigma^x ( -i \partial_x + \Xi_x)\\ &+\frac{\sqrt{1+ a_x^2}}{ \sqrt{a^2 +a^2_\theta + a^2_x a^2}}\sigma^y ( -i \partial_\theta + \Xi_{\theta} -i \frac{a_\theta a_x\partial_x}{ (1+a_x^2)} ) \notag
    \label{Hcurv2_main}
\end{align}
where
\begin{eqnarray}
    \Xi_x &=&  -i \frac{a_{\theta} a_{x \theta} (a_x^2 -3) -a_x (1+a_x^2)(2a + a_{\theta\theta})}{ 4(1+a_x^2 ) \sqrt{a^2 + a^2_\theta + a^2_x a^2}}, \\
    \Xi_\theta &=&  -i\frac{a_{xx} a_{\theta} (a^2_x -3) - a_{x\theta} a_x (1+a_x^2)}{4(1+a_x^2) \sqrt{a^2+ a^2_\theta + a_x^2 a^2}} + \frac{Ba^2}{2}
\end{eqnarray}
with $a_x = \frac{\partial a}{\partial x}$ and $a_\theta = \frac{\partial a}{\partial \theta}$. 
This Hamiltonian preserves the chiral symmetry at the Dirac point and therefore the perfectly transmitted mode remains.

\section{Discussion and conclusions}
In this work, we have explored the robustness of the perfectly transmitted mode, obtained at half a flux quantum through uniform topological insulator nanowires, to ripples in the surface.
We found that perfect transmission is generally not obtained, when the ripples are combined with either scalar disorder or finite doping (or both).
However, a chiral symmetry at the Dirac point in clean wires, is not broken by general surface deformations, and results in a $\mathbb{Z}$ topological classification of the wires.
A magnetic flux along the wire can tune between distinct topological phases, and we analytically calculated the flux value at which a transition between them is obtained (see Eq.~\eqref{flux_max}) in the case of azimuthally symmetric deformations.
Since the phases are characterized by a topological quantum number that is the sum of negative reflection eigenvalues, a perfect transmission is obtained at the transition points.
The combination of ripples and scalar disorder alter transport by locally breaking time-reversal symmetry.
In particular, as one of our main results, the amplitude of Aharonov-Bohm oscillations is found to decrease with increasing magnetic flux and eventually, for large ripples, vanishes.
Our findings may be important in carefully analyzing and understanding transport experiments on TI nanowires: our mechanism contributes to the loss of Aharonov-Bohm oscillations at large magnetic fields observed in magnetoresistance experiments~\cite{Cho2015,Dufouleur2017,Jauregui2016a,Hong2014,Dufouleur2017}; the effective Hamiltonians we derive also simplify the theoretical analysis of magnetotransport experiments with systematically nonuniform cross sections~\cite{Ziegler2018,Moors2018}.

A natural question raised by our work, is what happens to the Majorana mode obtained in TI nanowires when proximitized by a superconductor.
In order to obtain topological superconductivity the odd number of modes, related to the perfectly transmitted mode, are important. 
Since ripples generally break the protection of the perfectly transmitted mode when combined with scalar disorder, it is an important future research direction to understand in detail the interplay of ripples and superconductivity; however, our results suggest that at low chemical potentials and not too large ripple sizes, transport still resolves odd number of modes in the presence of a flux quantum. When coupled to superconductiviy, this suggests that the Majorana mode will still be present.

\acknowledgements
This work was supported by the ERC Starting Grant No.~679722 and the Knut and Alice Wallenberg Foundation 2013-0093. J.-W. Rhim was supported by the Research Center Program of the Institute for Basic Science in Korea No. IBS-R009-D1.

\appendix

\section{The Dirac equation for nanowires with a rippled surface }
\label{sec:DiracEquation}
In this section, we derive the Dirac Hamiltonian describing the surface states of a nanowire surface with smooth deformations of the cylindrical surface.
We provide two complementary derivations. 
First, we consider a two dimensional field theoretical approach that describes fermions on a curved manifold. 
We then revisit the problem from the lattice point of view\cite{Takane} where we construct an effective surface Hamiltonian starting from a microscopic Hamiltonian of the topological insulator.

\subsection{A field theoretic derivation}
The field theoretical description of relativistic particles on curved spacetime background is based on the fundamental principles of Lorentz invariance and general covariance.
Based on these symmetry principles, spin one-half particles, placed on a $2+1$ dimensional curved manifold with metric
\begin{eqnarray}
	ds^{2}=&g_{\mu\nu}(x) dx^{\mu} dx^{\nu} \nonumber \\
    = &V_{\mu}^{\alpha}(x)\eta_{\alpha\beta}V_{\nu}^{\beta}(x)dx^{\mu} dx^{\nu},
	\label{metric-vielbein}
\end{eqnarray}
are described by the covariant Dirac equation
\begin{equation}
	\gamma^{\mu}\nabla_{\mu} \psi =
    \gamma^{\mu} (\partial_{\mu} + \Gamma_{\mu})\psi =  0.
	\label{Dirac}
\end{equation}
Here, the spin connection term $\Gamma_{\mu}$---which describes the rotation of the spinors in a local coordinate frame---is a linear combination of the Dirac matrices with metric dependent coefficients \cite{Utiyama1956},
\begin{eqnarray}
	\Gamma_{\mu}(x) =&\frac{1}{2} \Sigma^{\alpha\beta} V_{\alpha}^{\nu}(x)(\partial_{\mu}V_{\beta\nu}-
	\Gamma_{\mu\nu}^{\rho}V_{\beta\rho}), \nonumber \\
	\Gamma_{\mu\nu}^{\rho} =& \frac{1}{2}g^{\rho\sigma}(\partial_{\mu}g_{\sigma\nu}+ \partial_{\nu}g_{\sigma\mu} - \partial_{\sigma}g_{\mu\nu}),
	\label{covariant}
\end{eqnarray}
with $\Gamma^{\rho}_{\mu\nu}$ the Christoffel symbols.
To make the notation flow better, we indicate with the first eight greek letters the Minkowskian metric indices, but with the last sixteen the curved spacetime metric indices; repeated indices are summed over.
The vielbeins $V_{\mu}^{\alpha}(x)$ are the tangent vectors at each point $X$ with normal coordinates $y^{\alpha}$ on the manifold 
\begin{equation}
    V_{\mu}^{\alpha}(X) = \left(\frac{\partial y^{\alpha}_{X}}{ \partial x ^{\mu}}\right)_{x=X}
\end{equation}
The generators of the Lorentz group, $\Sigma^{\alpha\beta}$, are given in terms of the Dirac matrices $\xi^{\alpha}$, which satisfy the Clifford algebra $ \{ \xi^{\alpha},\xi^{\beta} \}= 2\eta^{\alpha\beta}$, by 
\begin{equation}
\Sigma^{\alpha\beta} = \frac{1}{2} \left[\xi^{\alpha}, \xi^{\beta}\right].
    \label{LorGen}
\end{equation}
By definition, the local Dirac matrices $\xi^{\alpha}$ do not transform under a general coordinate transformation, in contrast to their covariant Dirac matrix counterparts
\begin{eqnarray}
\gamma^{\mu}(x) & = & V_{\alpha}^{\mu}(x) \xi^{\alpha}\label{DirSpin}.
\end{eqnarray}

A convenient representation of the Dirac matrices $\xi^{\alpha}$ in two spatial dimension is
\begin{equation}
	\xi^0 = -i \sigma^{z}, \quad\xi^1 = \sigma^y, \quad\xi^2 = -\sigma^x,
\end{equation}
such that the generators of the Lorentz group \eqref{LorGen} are written in the form
\begin{equation}
	\Sigma^{01} = - \frac{1}{2} \sigma^x, \quad \Sigma^{02} = -\frac{1}{2}\sigma^y, \quad \Sigma^{12} = \frac{i}{2} \sigma^z.
    \label{SpinConnection}
\end{equation}
Since the metric is independent of time, there is an associated Killing vector in the time direction indicating time translation invariance of the equation of motion and therefore an associated conserved quantity: the energy.
We can thus define a Hamiltonian, and from Eq.~(\ref{Dirac}) we have
\begin{eqnarray}
	i\partial_0 \psi &=&  H \psi, \nn 
    H &=&   i \gamma^0 \gamma^1 (\partial_1 +\Gamma_1)+ i \gamma^0 \gamma^2(\partial_2 + \Gamma_2).
    \label{Hcurv1}
\end{eqnarray}

The general metric for the surface is written in the form
\begin{equation}
    ds^2 = (1+ a_x^2) dx^2 + 2 a_x a_{\theta} dx d\theta +(a^2 +a^2_{\theta})d\theta^2 - dt^2
\end{equation}
where $a_x = \frac{\partial a}{\partial x}$ and $a_{\theta} = \frac{\partial a}{\partial \theta}$.
In the matrix form the metric $g_{\mu\nu}$ and the inverse metric $g^{\mu\nu}$ are written as

\begin{eqnarray}
    g_{\mu\nu} =
    \begin{bmatrix}
        -1 & 0 & 0 \\
        0 &  1 + a_x^2 & a_x a_{\theta} \\
        0 & a_x a_{\theta} & a^2 + a^{2}_{\theta}
    \end{bmatrix},
\end{eqnarray}
\begin{eqnarray}
    g^{\mu\nu} = \frac{1}{\mathrm{det}g}
    \begin{bmatrix}
        -1 & 0 & 0 \\
        0  &  a^2 + a^{2}_{\theta} & - a_x a_{\theta} \\
        0  & -a_x a_{\theta}       & 1 + a_x^2
    \end{bmatrix}
\end{eqnarray}
where $\mathrm{det} g = a^2 + a_{\theta}^2 +a_x^2 a^2$

If we fix the gauge, the vielbein matrix is written in the form

\begin{eqnarray}
    V_{\mu }^{\alpha}  = 
    \begin{bmatrix}
        1 & 0 & 0 \\
        0 &  \sqrt{1 + a_x^2} & 0  \\
        0 & \frac{a_x a_{\theta} }{\sqrt{1 + a_x^2}}& \sqrt{\frac{\mathrm{det}g}{1+a_x^2}}
    \end{bmatrix},
\end{eqnarray}
that satisfies Eq.~\eqref{metric-vielbein}. 
Note again that $\mu$ refers to the curved space index, and $\alpha$ to the local coordinate system.
We can transform these indices from covariant to contravariant by using $g_{\mu\nu}$ and $\eta_{\alpha\beta}$ respectively. 
In that case the inverse vielbein matrix is 
\begin{equation}
    V^\mu_\alpha =
   \begin{bmatrix}
        1 & 0 & 0 \\
 0 & \frac{1}{\sqrt{a_x^2+1}} & -\frac{a_\theta a_x}{\sqrt{a_x^2+1} \sqrt{det g}} \\
 0 & 0 & \frac{\sqrt{a_x^2+1}}{\sqrt{det g}} \\
   \end{bmatrix}.
    \label{}
\end{equation}

Christoffel symbols with at least one index zero vanish; the remaining ones are
\begin{eqnarray}
    \Gamma^1_{11} &=& \frac{1}{ det g} a_x a_{xx} a^2,\\
    \Gamma^1_{12} &=& \frac{1}{ det g} a a_x (a a _{x\theta} - a_x a_\theta) = \Gamma^1_{21},\\
    \Gamma^1_{22} &=& \frac{1}{ det g} a a_x (a a_{\theta\theta} - a^2 -2a^2_{\theta}), \\
    \Gamma^2_{11} &=& \frac{1}{ det g} a_{xx} a_{\theta},\\
    \Gamma^2_{12} &=& \frac{1}{ det g} a a_x (1+ a_x^2) a_{\theta} a_{x \theta} = \Gamma^2_{21},\\
    \Gamma^2_{22} &=& \frac{1}{ det g} a_{\theta  }( 2a_{x}^2 a + a +a_{\theta\theta}).
    \label{ChristSymb}
\end{eqnarray}
The spin connections $\Gamma_1=\Gamma_x$ is written in the form 
\begin{eqnarray}
	\Gamma_{x} &=&\frac{1}{2} \Sigma^{\alpha\beta} V^{\nu}_{\alpha}(\partial_{x}V_{\nu\beta}-
	\Gamma_{x\nu}^{\rho}V_{\rho\beta}) \\
	&=&\frac{1}{2} \Sigma^{01} V_{0}^{\nu}(\partial_{x}V_{\nu 1}-
	\Gamma_{x\nu}^{\rho}V_{\rho 1}) \nonumber\\
	&&+\frac{1}{2} \Sigma^{02} V_{0}^{\nu}(\partial_{x}V_{\nu 2}-
	\Gamma_{x\nu}^{\rho}V_{\rho 2}) \nonumber\\
	&&+\frac{1}{2} \Sigma^{10} V_{1}^{\nu}(\partial_{x}V_{\nu 0}-
	\Gamma_{x\nu}^{\rho}V_{\rho 0}) \nonumber\\
	&&+\frac{1}{2} \Sigma^{12} V_{1}^{\nu}(\partial_{x}V_{\nu 2}-
	\Gamma_{x\nu}^{\rho}V_{\rho 2}) \nonumber\\
	&& + \frac{1}{2} \Sigma^{20} V_{2}^{\nu}(\partial_{x}V_{\nu 0}-
	\Gamma_{x\nu}^{\rho}V_{\rho 0}) \nonumber \\
	&& + \frac{1}{2} \Sigma^{21} V_{2}^{\nu}(\partial_{x}V_{\nu 1}-
	\Gamma_{x\nu}^{\rho}V_{\rho 1}).
\end{eqnarray}
By doing a straightforward calculation we find that the coefficients of $\Sigma^{01} , \Sigma^{02}$ vanish and $\Gamma_x$ is simplified to the form
\begin{eqnarray}
	\Gamma_x &=&\frac{1}{2} \Sigma^{12} V_{1}^{\nu}(\partial_{x}V_{\nu 2}-
	\Gamma_{x\nu}^{\rho}V_{\rho 2}) \nonumber\\
	 &&+ \frac{1}{2} \Sigma^{21} V_{2}^{\nu}(\partial_{x}V_{\nu 1}-
	\Gamma_{x\nu}^{\rho}V_{\rho 1}). 
\end{eqnarray}
By calculating the coefficients, we find that the spin connection in the $x$ direction is given by the form
\begin{eqnarray}
     \Gamma_x = \Sigma^{12}G_x(x,\theta), \\
     \Gamma_x = \frac{i}{4} \sigma^z G_x(x,\theta),
\end{eqnarray}
where the function $G_x$ is given by
\begin{eqnarray}
    G_x = \frac{a_{xx} a_{\theta} (a^2_x -3) - a_{x\theta} a_x (1+a_x^2)}{(1+a_x^2) \sqrt{g}}.
\end{eqnarray}
By a similar calculation, we find for $\Gamma_{\theta}$
\begin{eqnarray}
	\Gamma_\theta =\frac{1}{2} \Sigma^{12} V_{1}^{\nu}(\partial_{\theta}V_{\nu 2}-
	\Gamma_{\theta\nu}^{\rho}V_{\rho 2}) \nonumber\\
	 + \frac{1}{2} \Sigma^{21} V_{2}^{\nu}(\partial_{\theta}V_{\nu 1}-
	\Gamma_{\theta\nu}^{\rho}V_{\rho 1}),
\end{eqnarray}
that is written as
\begin{equation}
    \Gamma_\theta = \frac{i}{4} \sigma^z G_\theta(x,\theta),
\end{equation}
so that 
\begin{equation}
    G_{\theta } = \frac{a_{\theta} a_{x \theta} (a_x^2 -3) -a_x (1+a_x^2)(2a + a_{\theta\theta})}{ (1+a_x^2 ) \sqrt{\mathrm{det}g}}.
\end{equation}
We notice also that 
\begin{eqnarray}
    i\gamma^0 \gamma^1 &=&  -i V_1^1\sigma^x -i V_2^1\sigma^y \nn &=&  -i \frac{1}{ \sqrt{1+a_x^2}} \sigma^x -i\frac{a_\theta a_x}{ \sqrt{\mathrm{det} g (1+a_x^2)}} \sigma^y,
\end{eqnarray}
and
\begin{equation}
    i\gamma^0 \gamma^2 =  -i V_2^1\sigma^y = -i \frac{\sqrt{1+ a_x^2}}{ \sqrt{\mathrm{det} g}} \sigma^y.
\end{equation}
Putting all together, we find that Eq.~\eqref{Hcurv1} is written as
\begin{eqnarray}
    H_{\Xi} &=& \frac{1}{ \sqrt{1+a_x^2}} \sigma^x ( -i \partial_x + \Xi_x)\nonumber\\ &&+\frac{\sqrt{1+ a_x^2}}{ \sqrt{a^2 +a^2_\theta + a^2_x a^2}}\sigma^y ( -i \partial_\theta + \Xi_{\theta} -i \frac{a_\theta a_x\partial_x}{ (1+a_x^2)}  ),\nonumber\\
    \label{Hcurv2}
\end{eqnarray}
where
\begin{eqnarray}
    \Xi_x &=&  -i \frac{a_{\theta} a_{x \theta} (a_x^2 -3) -a_x (1+a_x^2)(2a + a_{\theta\theta})}{ 4(1+a_x^2 ) \sqrt{a^2 + a^2_\theta + a^2_x a^2}}, \\
    \Xi_\theta &=&  -i\frac{a_{xx} a_{\theta} (a^2_x -3) - a_{x\theta} a_x (1+a_x^2)}{4(1+a_x^2) \sqrt{a^2+ a^2_\theta + a_x^2 a^2}}. 
\end{eqnarray}
For the geometry we use in the main part of this paper, we choose the coordinate system to be
\begin{equation}
	g_{\mu\nu} = \mathrm{diag}[-1, 1+\partial_x a^2(x), a^2(x)].
    \label{metric_ripple}
\end{equation}
In that case $a_\theta =0 $ and from Eq.~\eqref{Hcurv2} we obtain the Dirac Hamiltonian for a rippled wire
\begin{eqnarray}
	H &=&   -i\frac{1}{\sqrt{1+(\partial_x a)^2}}\sigma^x \partial_x -\frac{i}{a(x)}\sigma^y \partial_{\theta}
     \nn
    &&- \frac{i\partial_x a}{2a(x)\sqrt{1+(\partial_x a)^2}} \sigma^x.
\end{eqnarray}

In the above derivation we had used the system of natural units.
By restoring the fundamental constants and coupling the magnetic flux minimally to the azimuthal momentum
\begin{equation}
    -i \hbar  \partial_{\theta} \rightarrow
    -i \hbar  \partial_{\theta} +e A_{\theta},
    \label{}
\end{equation}
we obtain the Hamiltonian that appears in the main text 
\begin{eqnarray}
    H &=& v_F [\sigma_x g_{xx}^{-1/2} (-i\hbar\partial_x + A_x) \nn
    &&+ \sigma_yg_{\theta\theta}^{-1/2} (-i \hbar\partial_{\theta} + e A_\theta) ].
	\label{Dirac_fluc_rad2}
\end{eqnarray}
This equation implies general covariance of the model. 
While general covariance is a fundamental symmetry in relativistic system, in a condensed matter systems it is emergent.
For a three dimensional topological insulator, it comes together with the Dirac nature of the surface fermions.
In the next section we provide a microscopic derivation of Eq.~\eqref{Dirac_fluc_rad2} that does not rely on covariance but rather on a three dimensional, low energy effective Hamiltonian for a lattice model of a three dimensional topological insulator.

\subsection{A microscopic derivation}
In this subsection we derive \eqref{Dirac_fluc_rad2} from the microscopic model of Takane et al. \cite{Takane}.
We start from a low energy effective Hamiltonian for the bulk of the three dimensional topological insulator, 
\begin{equation}
H_\mathrm{bulk} = (m_0 + m_2 \v p^2)\tau^z +  \tau^x (\boldsymbol\sigma\cdot\v p),
    \label{}
\end{equation}
where $\v p = -i\nabla$, $\tau^\alpha$ and $\sigma^\alpha$ are Pauli matrices for the real spin and orbital degrees of the freedom, and $m_0, m_2$ are free parameters.
The surface of the system is described by the two-parameter-dependent three-dimensional normal vector $\v X(x^1,x^2)$.
The comoving coordinate system in the surface has the basis vectors
\begin{equation}
    \v e_i = \partial \v X/\partial x^i
    \label{}
\end{equation}
for $i=1,2,$ and the normal component $\v e_3 = \v e^3 = \v e_1 \times \v e_2/ |\v e_1 \times \v e_2|$, which together define a three dimensional metric for the surface
\begin{equation} 
    g^{3D}_{ij} = \v e_i\cdot \v e_j.
    \label{metric3d}
\end{equation}
By assuming that the wave function is localized at the boundary Takane et al.~\cite{Takane} derived the effective surface Hamiltonian
\be
H_{\mathrm{eff}} = \bpm 0 & \mathcal{D}_+ \\ \mathcal{D}_- & 0 \epm,
\label{Heff}
\ee
where
\begin{widetext}
\be
\mathcal{D}_+ &=& \sum_{i=1}^2 \left\{ \left( \eta_i -m_2\xi_i \right)\left( \partial_i + \frac{1}{2}\left[ \partial_i \ln\langle \sqrt{g}\rangle \right] \right) +\frac{1}{2}\left[ \partial_i \left( \eta_i -m_2\xi_i \right) \right] \right\}, \label{eq:D_+}\\
\mathcal{D}_- &=& \sum_{i=1}^2 \left\{ -\left( \eta_i -m_2\xi_i \right)^*\left( \partial_i + \frac{1}{2}\left[ \partial_i \ln\langle \sqrt{g}\rangle \right] \right) -\frac{1}{2}\left[ \partial_i \left( \eta_i -m_2\xi_i \right)^* \right] \right\}.
\label{eq:D_-}
\ee
\end{widetext}
Here
\begin{align}
\eta_i &= \frac{\langle \sqrt{g} \v n^\dag_+ \sigma^i \v n_-\rangle}{\langle \sqrt{g} \rangle}, \\
 \xi_i &= 2\sum_{j=1}^2 \frac{\langle \sqrt{g} g^{ij} \v n^\dag_+ \partial_j \v n_-\rangle}{\langle \sqrt{g} \rangle},
\end{align}
where $\sigma^i = \v e^i\cdot\boldsymbol\sigma$ and $\boldsymbol \sigma = (\sigma^x, \sigma^y, \sigma^z)$.
The inverse metric $g^{ij} = \v e^i\cdot \v e^j$ is obtained by the condition $\v e_i\cdot\v e^j = \delta_{ij}$.
The two spinors $\v n_+$ and $\v n_-$ are the eigenvectors of $\sigma^3$ satisfying $\sigma^3\v n_\pm = \pm\v n_\pm$.

For a rippled nanowire the vector that describes the surface is
\be
\v X(x^1,x^2) = \v X(\varphi,z) = \left[ a(z)\cos\varphi, a(z)\sin\varphi, z \right]
 \label{eq:Xsurf}
\ee
in polar coordinates $(r,\varphi, z)$.
By plugging \eqref{eq:Xsurf} into \eqref{metric3d} we find that Eq.\eqref{Heff} takes the form
\begin{widetext}
\begin{eqnarray}
H_{\mathrm{eff}} &=& -i[ 1+\frac{m_2}{a\sqrt{1+(\partial_z a)^2}}]\frac{1}{a}\sigma^x\partial_\varphi 
\nn
&-&i [ \frac{1}{\sqrt{1+(\partial_z a)^2}} 
- \frac{m_2 \partial^2_z a}{(1+(\partial_z a)^2)^2}]\sigma^y\partial_z 
-i [\frac{\partial_z a}{2a\sqrt{1+(\partial_z a)^2}} 
+ m_2 F(z)]\sigma^y ,
\label{eq:H_eff}
\end{eqnarray}
where
\be
F(z) = -\frac{\partial_z a \partial^2_z a}{2 a [1+(\partial_z a)^2]^2} 
+3 \frac{ \partial_z a (\partial^2_z a)^2}{2[ 1+(\partial_z a)^2]^3} 
- \frac{\partial_z^3 a}{2[ 1+(\partial_z a )^2 ]^2} .
\ee
\end{widetext}
Here, we drop the argument $z$ in $a(z)$ for convenience.
We consider the isotropic case ($m_2=0$) where the effective Hamiltonian reduces to
\begin{eqnarray}
H_{\mathrm{eff}} &=&  -i \frac{1}{a}\sigma^x \partial_\varphi -i \frac{1}{\sqrt{1+(\partial_z a)^2}}\sigma^y\partial_z 
\nn
&-& i\frac{\partial_z a}{2a\sqrt{1+(\partial_z a)^2}}\sigma^y  .
\label{eq:H_eff_iso}
\end{eqnarray}
If we couple the electromagnetic potential, substitute the coordinate $z$ with $x$ and $\varphi$ with $\theta$, restore $\hbar, v_F$ and perform the unitary transformation of the Hamiltonian by
$U = i (\sigma^x + \sigma^y)/\sqrt{2}$, we arrive at the equation found in the main text.
\section{The transfer matrix method for nanowires with a rippled surface}
\label{sec:Transfer}
In this section we discuss the details of the transfer matrix method \cite{Mello1988,Beenakker1997,Bardarson2013} adapted to a rippled nanowire and used in the main text.
In a multichannel conductor, the transport mode basis state $\ket{n,\pm}$ is an eigenstate of the current operator $\Sigma^z_{nm} = \sigma^z \delta_{nm}$ with eigenvalue $\pm$, and therefore, a scattering state
\begin{equation}
    \ket{\psi} = \sum_n C_n^{+} \ket{n,+}  + C_{n}^{-}\ket{n,-}
\end{equation}
carries a total current 
\cite{Mello1988},
\begin{equation}
    I  =  \sum_n (|C_n^{+}|^2 -|C_n^{-}|^2).
\end{equation}
For the topological insulator nanowire, the unitary transformation $\psi \rightarrow R \psi$, where $R = \frac{1}{\sqrt{2}} \bigl(\begin{smallmatrix}1 &1 \\ 1 &-1\end{smallmatrix} \bigr)$ rotates the current operator $j_x = \frac{\partial H}{\partial p_x}$ to $\Sigma^z$; for a uniform wire with radius $a_0$ the scattering state with momentum $k$ at high energy---appropriate for a state modeling a lead---takes the form 
\begin{eqnarray}
\langle x,\theta | n,+ \rangle &=&  \frac{1}{\sqrt{2 \pi a_0}}e^{ +i\theta(n-\frac{1}{2})}e^{ikx}\,\,\, \begin{pmatrix} 1 \\ 0 \end{pmatrix}, \nn
    \langle x,\theta | n,- \rangle &=&  \frac{1}{\sqrt{2 \pi a_0}}e^{ +i\theta(n-\frac{1}{2})}e^{-ikx} \begin{pmatrix} 0 \\ 1 \end{pmatrix}.
    \label{transport_mode}
\end{eqnarray}
The states are normalized according to the inner product
\begin{equation}
    \langle \psi | \psi \rangle = \int_{0}^{2\pi} a_0 d\theta  \psi^{\dagger}(x,\theta)\psi(x,\theta).
    \label{}
\end{equation}
If the radius fluctuates, $a = a(x)$, 
the wave function is
\begin{eqnarray}
	\psi(x, \theta) &=& \sum_n \Bigg[ C^{+}_n \frac{1}{\sqrt{2 \pi a(x)}}e^{ +i\theta(n-\frac{1}{2})}e^{ikx} \begin{pmatrix} 1 \\ 0 \end{pmatrix}
        \nn
		&&+ C^{-}_n \frac{1}{\sqrt{2\pi a(x)}}e^{ i\theta(n-\frac{1}{2})}e^{-ikx} \begin{pmatrix} 0 \\ 1 \end{pmatrix} \Bigg]. \nonumber \\
        \label{scatstate}
\end{eqnarray}

The Schr{\"o}dinger equation $H\psi = E_F \psi$ of \eqref{Dirac} for a transport mode of a disordered wire, 
\begin{eqnarray}
	\phi_n(x) &=&  \begin{pmatrix} m_{1,n}(x) \\ m_{2,n}(x) \end{pmatrix}  
    \nn
    &=&  \int_{0}^{2\pi} d\theta e^{-i\theta(n-\frac{1}{2})}\sqrt{2\pi a(x)} \psi(x, \theta),
	\label{scat-states}
\end{eqnarray}
is written in the form
\begin{equation}
    \partial_x \phi_n(x) = M_{nm}(x) \phi_m(x),
\end{equation}
where
\begin{eqnarray}
	M_{m,n}(x) & = & \frac{i}{\hbar v_F}\left[ E_F \delta_{n,m} - V_{nm}(x) \right] \sqrt{1+(\partial_x a)^2} \sigma^z\nonumber \\
    &&+ \delta_{n,m}\frac{\sqrt{1+(\partial_x a)^2}}{a}[n -\frac{1}{2} +e\frac{B a^2}{2\hbar}]\sigma^x.
	\label{transfermatrix_appendix}
\end{eqnarray}
Thus, we obtain the transfer matrix of the spinor that appears in the main text, by integrating 
\begin{eqnarray}
	T_{m,n}& = &\mathcal{P}_{x_s}  \exp\left[\int_{0}^{L}dx_s\ M_{m,n}(x_s) \right],
\end{eqnarray}
where $\mathcal{P}_{x_s}$ is the position ordering operator and $V_{nm}= \frac{1}{2 \pi} \int_{0}^{2\pi} d\theta e^{i(n-m)\theta}V(x, \theta)$.

The transfer matrix is related to the scattering matrix elements by the relation \cite{Beenakker1997}
\begin{eqnarray}
	T~=~\begin{pmatrix} t^{\dagger -1}&r^{\prime} t^{\prime -1} \\
		- t^{\prime -1} r&t^{\prime -1} \end{pmatrix},
	\label{trans-scat}
\end{eqnarray}
and in general, is hard to compute because of the position ordering operator. 
However, at the special case of zero disorder $K_V = 0$ and at the Dirac point $E_F=0$, we can evaluate this matrix to find

\be
T_{m,n} &=& \bpm \cosh\theta_m  & -i\sinh\theta_m \\ i\sinh\theta_m & \cosh\theta_m \epm,
\ee
where 
\be
\theta_m = \int_0^L dx a^{-1}\sqrt{1+(\partial_x a)^2}\left(m -\frac{1}{2} +\frac{e}{2\hbar}B a^2\right),
\label{eq:thetam}
\ee 
yielding
\be
t^\prime = t = \sech\theta_m, \label{eq:t}
\ee
and
\be
r^\prime = r = -i\tanh\theta_m. \label{eq:s}
\ee
We obtain a perfectly transmitted mode when $\theta_m=0$, or when
\begin{equation}
    \frac{\phi_{\textrm{max},n}}{\phi_0} = a_0^2\left(n+ \frac{1}{2}\right)\frac{  \int_{0}^{L} dx~ a^{-1} \sqrt{1 + (\partial_x a)^2}}{\int_{0}^{L} dx~ a \sqrt{1 + (\partial_x a)^2}},
\end{equation} 
as given in in the main text.

\section{Disorder averaged conductance and perfect transmission}
\label{sec:DisorderAverage}
\begin{figure}[tb!]
    \centering
    \includegraphics[width=\columnwidth]{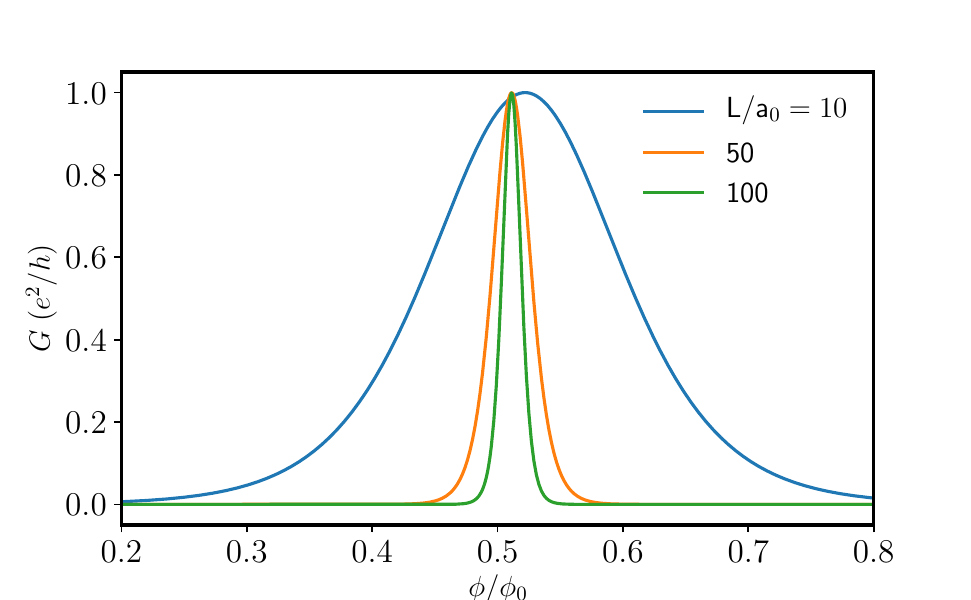} 
	\includegraphics[width=\columnwidth]{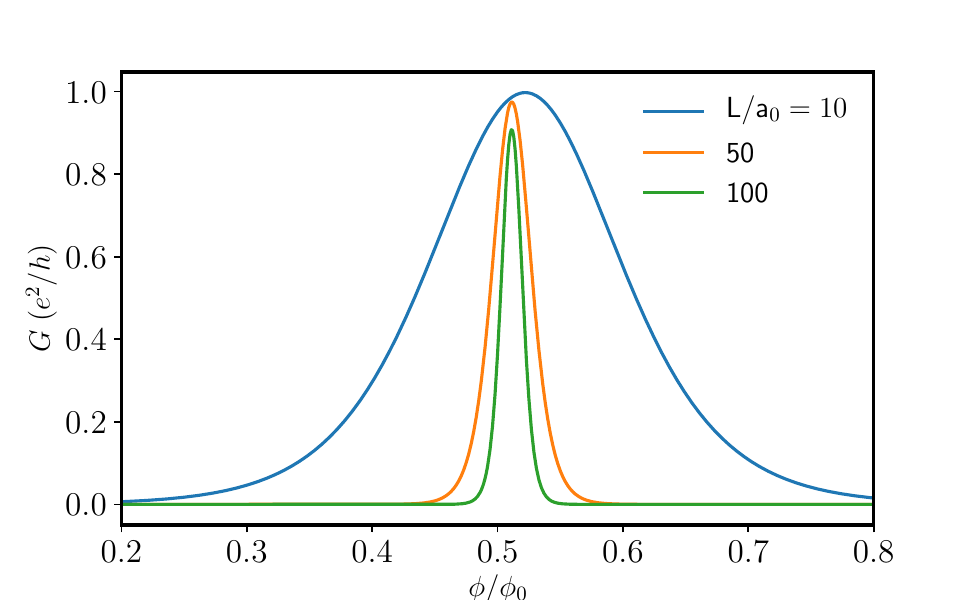}
    \caption{Conductance as function of magnetic flux for a wire with a varying radius given by $a(x) = a_0 + \delta a \cos(\lambda x)$, where $\delta a \in [0.1a_0,0.3a_0]$ and $\lambda = 1/a_0$. The top panel shows the conductance for a given realization ($\delta a = 0.2$) and the lower panel shows the disorder average.}
    \label{fig:ptmAv}
\end{figure}
As argued in the main text, perfect transmission is obtained in any given fixed disorder realization.
This seemingly contradicts theories of the localization transition that predict conductance at the critical point that decays to zero as system size goes to infinity, $\langle G \rangle \sim L^{-1/2}$.\cite{balents1997delocalization,brouwer2000localization,motrunich2001griffiths,gruzberg2005localization,Evers2008,altland2015topology}
However, this contradiction is only apparent, and arises from the fact that the value of the critical flux $\phi_c$ for a \emph{given} disorder realization, depends on the realization. 
When one calculates the average conductance, this is done at a \emph{fixed} flux $\phi$ which is taken to be equal to the average of the critical flux for the disorder ensemble, $\phi = \langle \phi_c \rangle$. 
This will give rise to non-quantized conductance as demonstrated in Fig.~\ref{fig:ptmAv}.
Here we calculate the conductance for a simplified model of the random variation of the radius, with $a(x) = a_0 + \delta a \cos(\lambda x)$ with $\delta a$ uniformly distributed.
The conductance can be obtained by substituting this form into the expression~\eqref{eq:thetam} for $\theta_m$, that in turns gives the conductance via Eq.~\eqref{eq:t}.

\begin{figure}[tb!]
    \centering
	\includegraphics[width=\columnwidth]{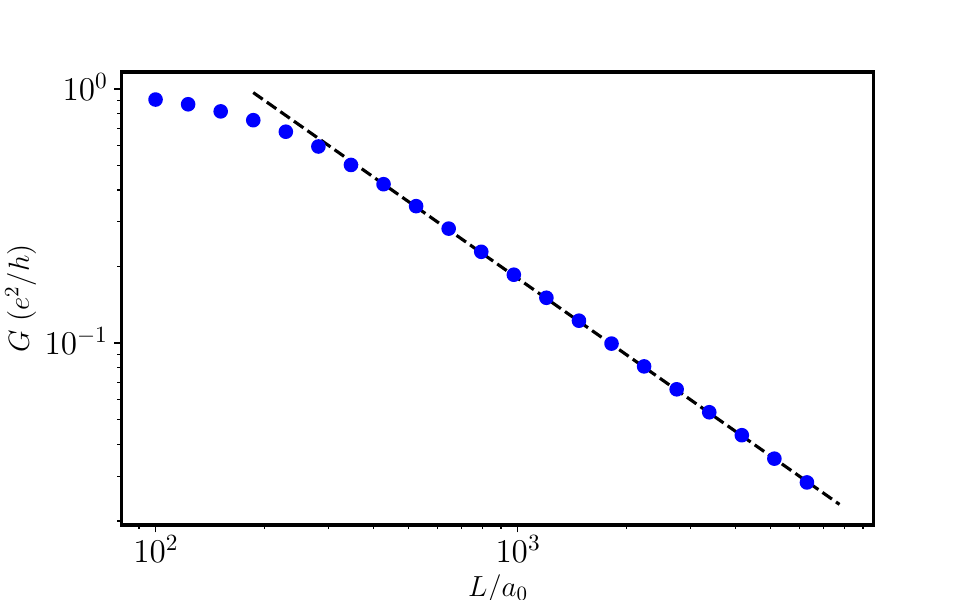}
    \caption{Conductance as a function of system size $L$ for the model and parameters defined in Fig.~\ref{fig:ptmAv} at a fixed $\phi = 0.511$. The dashed lines is a guide to the eye and presents decay going like $L^{-1}$. }
    \label{fig:gvL}
\end{figure}

The decrease of the disorder averaged conductance with system size is shown in Fig.~\ref{fig:gvL}.
Over a large range of systems sizes, the conductance goes as $\langle G \rangle \sim 1/L$.
The $1/L$ dependence can be explained from the following considerations: in our simplified model used for these figures, the effect of the disorder is essentially to shift the critical flux strength to a different value of the flux, while the shape of the conductance as a function of flux remains essentially the same. 
From Eq.~\eqref{eq:thetam} we can model this as having $\theta_m = \delta \theta L$ with $\delta\theta$ uniformly distributed around zero.
The average conductance therefore will take the form
\be
\langle G \rangle = \frac{e^2}{h}\int_{-\theta_a}^{\theta_a} d\delta \theta \frac{1}{\cosh^2(\delta\theta L)} = \frac{2\tanh (\theta_a L)}{L},
\ee
consistent with what we observe.
This explanation further demonstrates that the exact decay of the average conductance obtained at the critical point will depend on the details of the disorder distribution used---we leave it for future research to determine this detailed relationship.

\bibliography{refs}

\begin{thebibliography}{51}%
\makeatletter
\providecommand \@ifxundefined [1]{%
 \@ifx{#1\undefined}
}%
\providecommand \@ifnum [1]{%
 \ifnum #1\expandafter \@firstoftwo
 \else \expandafter \@secondoftwo
 \fi
}%
\providecommand \@ifx [1]{%
 \ifx #1\expandafter \@firstoftwo
 \else \expandafter \@secondoftwo
 \fi
}%
\providecommand \natexlab [1]{#1}%
\providecommand \enquote  [1]{``#1''}%
\providecommand \bibnamefont  [1]{#1}%
\providecommand \bibfnamefont [1]{#1}%
\providecommand \citenamefont [1]{#1}%
\providecommand \href@noop [0]{\@secondoftwo}%
\providecommand \href [0]{\begingroup \@sanitize@url \@href}%
\providecommand \@href[1]{\@@startlink{#1}\@@href}%
\providecommand \@@href[1]{\endgroup#1\@@endlink}%
\providecommand \@sanitize@url [0]{\catcode `\\12\catcode `\$12\catcode
  `\&12\catcode `\#12\catcode `\^12\catcode `\_12\catcode `\%12\relax}%
\providecommand \@@startlink[1]{}%
\providecommand \@@endlink[0]{}%
\providecommand \url  [0]{\begingroup\@sanitize@url \@url }%
\providecommand \@url [1]{\endgroup\@href {#1}{\urlprefix }}%
\providecommand \urlprefix  [0]{URL }%
\providecommand \Eprint [0]{\href }%
\providecommand \doibase [0]{http://dx.doi.org/}%
\providecommand \selectlanguage [0]{\@gobble}%
\providecommand \bibinfo  [0]{\@secondoftwo}%
\providecommand \bibfield  [0]{\@secondoftwo}%
\providecommand \translation [1]{[#1]}%
\providecommand \BibitemOpen [0]{}%
\providecommand \bibitemStop [0]{}%
\providecommand \bibitemNoStop [0]{.\EOS\space}%
\providecommand \EOS [0]{\spacefactor3000\relax}%
\providecommand \BibitemShut  [1]{\csname bibitem#1\endcsname}%
\let\auto@bib@innerbib\@empty
\bibitem [{\citenamefont {Moore}(2010)}]{Moore2010}%
  \BibitemOpen
  \bibfield  {author} {\bibinfo {author} {\bibfnamefont {Joel~E}\ \bibnamefont
  {Moore}},\ }\bibfield  {title} {\enquote {\bibinfo {title} {{The birth of
  topological insulators}},}\ }\href {\doibase 10.1038/nature08916} {\bibfield
  {journal} {\bibinfo  {journal} {Nature}\ }\textbf {\bibinfo {volume} {464}},\
  \bibinfo {pages} {194--198} (\bibinfo {year} {2010})}\BibitemShut {NoStop}%
\bibitem [{\citenamefont {Hasan}\ and\ \citenamefont {Kane}(2010)}]{Hasan2010}%
  \BibitemOpen
  \bibfield  {author} {\bibinfo {author} {\bibfnamefont {M.~Z.}\ \bibnamefont
  {Hasan}}\ and\ \bibinfo {author} {\bibfnamefont {C.~L.}\ \bibnamefont
  {Kane}},\ }\bibfield  {title} {\enquote {\bibinfo {title} {{Colloquium:
  Topological insulators}},}\ }\href {\doibase 10.1103/RevModPhys.82.3045}
  {\bibfield  {journal} {\bibinfo  {journal} {Rev. Mod. Phys.}\ }\textbf
  {\bibinfo {volume} {82}},\ \bibinfo {pages} {3045--3067} (\bibinfo {year}
  {2010})}\BibitemShut {NoStop}%
\bibitem [{\citenamefont {Qi}\ and\ \citenamefont {Zhang}(2011)}]{Qi2011}%
  \BibitemOpen
  \bibfield  {author} {\bibinfo {author} {\bibfnamefont {Xiao-Liang}\
  \bibnamefont {Qi}}\ and\ \bibinfo {author} {\bibfnamefont {Shou-Cheng}\
  \bibnamefont {Zhang}},\ }\bibfield  {title} {\enquote {\bibinfo {title}
  {{Topological insulators and superconductors}},}\ }\href {\doibase
  10.1103/RevModPhys.83.1057} {\bibfield  {journal} {\bibinfo  {journal} {Rev.
  Mod. Phys.}\ }\textbf {\bibinfo {volume} {83}},\ \bibinfo {pages}
  {1057--1110} (\bibinfo {year} {2011})}\BibitemShut {NoStop}%
\bibitem [{\citenamefont {Bardarson}\ and\ \citenamefont
  {Moore}(2013)}]{Bardarson2013}%
  \BibitemOpen
  \bibfield  {author} {\bibinfo {author} {\bibfnamefont {Jens~H.}\ \bibnamefont
  {Bardarson}}\ and\ \bibinfo {author} {\bibfnamefont {Joel~E.}\ \bibnamefont
  {Moore}},\ }\bibfield  {title} {\enquote {\bibinfo {title} {{Quantum
  interference and Aharonov–Bohm oscillations in topological insulators}},}\
  }\href {\doibase 10.1088/0034-4885/76/5/056501} {\bibfield  {journal}
  {\bibinfo  {journal} {Rep. Prog. Phys.}\ }\textbf {\bibinfo {volume} {76}},\
  \bibinfo {pages} {056501} (\bibinfo {year} {2013})}\BibitemShut {NoStop}%
\bibitem [{\citenamefont {Lee}(2009)}]{Lee2009}%
  \BibitemOpen
  \bibfield  {author} {\bibinfo {author} {\bibfnamefont {Dung-Hai}\
  \bibnamefont {Lee}},\ }\bibfield  {title} {\enquote {\bibinfo {title}
  {{Surface States of Topological Insulators: The Dirac Fermion in Curved
  Two-Dimensional Spaces}},}\ }\href {\doibase 10.1103/PhysRevLett.103.196804}
  {\bibfield  {journal} {\bibinfo  {journal} {Phys. Rev. Lett.}\ }\textbf
  {\bibinfo {volume} {103}},\ \bibinfo {pages} {196804} (\bibinfo {year}
  {2009})}\BibitemShut {NoStop}%
\bibitem [{\citenamefont {Zhang}\ and\ \citenamefont
  {Vishwanath}(2010)}]{Zhang2010}%
  \BibitemOpen
  \bibfield  {author} {\bibinfo {author} {\bibfnamefont {Yi}~\bibnamefont
  {Zhang}}\ and\ \bibinfo {author} {\bibfnamefont {Ashvin}\ \bibnamefont
  {Vishwanath}},\ }\bibfield  {title} {\enquote {\bibinfo {title} {{Anomalous
  Aharonov-Bohm conductance oscillations from topological insulator surface
  states}},}\ }\href {\doibase 10.1103/PhysRevLett.105.206601} {\bibfield
  {journal} {\bibinfo  {journal} {Phys. Rev. Lett.}\ }\textbf {\bibinfo
  {volume} {105}},\ \bibinfo {pages} {206601} (\bibinfo {year}
  {2010})}\BibitemShut {NoStop}%
\bibitem [{\citenamefont {Bardarson}\ \emph {et~al.}(2010)\citenamefont
  {Bardarson}, \citenamefont {Brouwer},\ and\ \citenamefont
  {Moore}}]{Bardarson2010}%
  \BibitemOpen
  \bibfield  {author} {\bibinfo {author} {\bibfnamefont {J.~H.}\ \bibnamefont
  {Bardarson}}, \bibinfo {author} {\bibfnamefont {P.~W.}\ \bibnamefont
  {Brouwer}}, \ and\ \bibinfo {author} {\bibfnamefont {J.~E.}\ \bibnamefont
  {Moore}},\ }\bibfield  {title} {\enquote {\bibinfo {title} {{Aharonov-Bohm
  oscillations in disordered topological insulator nanowires}},}\ }\href
  {\doibase 10.1103/PhysRevLett.105.156803} {\bibfield  {journal} {\bibinfo
  {journal} {Phys. Rev. Lett.}\ }\textbf {\bibinfo {volume} {105}},\ \bibinfo
  {pages} {156803} (\bibinfo {year} {2010})}\BibitemShut {NoStop}%
\bibitem [{\citenamefont {Ran}\ \emph {et~al.}(2008)\citenamefont {Ran},
  \citenamefont {Vishwanath},\ and\ \citenamefont {Lee}}]{Ran2008}%
  \BibitemOpen
  \bibfield  {author} {\bibinfo {author} {\bibfnamefont {Ying}\ \bibnamefont
  {Ran}}, \bibinfo {author} {\bibfnamefont {Ashvin}\ \bibnamefont
  {Vishwanath}}, \ and\ \bibinfo {author} {\bibfnamefont {Dung-Hai}\
  \bibnamefont {Lee}},\ }\bibfield  {title} {\enquote {\bibinfo {title}
  {{Spin-Charge Separated Solitons in a Topological Band Insulator}},}\ }\href
  {\doibase 10.1103/PhysRevLett.101.086801} {\bibfield  {journal} {\bibinfo
  {journal} {Phys. Rev. Lett.}\ }\textbf {\bibinfo {volume} {101}},\ \bibinfo
  {pages} {086801} (\bibinfo {year} {2008})}\BibitemShut {NoStop}%
\bibitem [{\citenamefont {Rosenberg}\ \emph {et~al.}(2010)\citenamefont
  {Rosenberg}, \citenamefont {Guo},\ and\ \citenamefont
  {Franz}}]{Rosenberg2010}%
  \BibitemOpen
  \bibfield  {author} {\bibinfo {author} {\bibfnamefont {G.}~\bibnamefont
  {Rosenberg}}, \bibinfo {author} {\bibfnamefont {H.-M.}\ \bibnamefont {Guo}},
  \ and\ \bibinfo {author} {\bibfnamefont {M.}~\bibnamefont {Franz}},\
  }\bibfield  {title} {\enquote {\bibinfo {title} {{Wormhole effect in a strong
  topological insulator}},}\ }\href {\doibase 10.1103/PhysRevB.82.041104}
  {\bibfield  {journal} {\bibinfo  {journal} {Phys. Rev. B}\ }\textbf {\bibinfo
  {volume} {82}},\ \bibinfo {pages} {041104(R)} (\bibinfo {year}
  {2010})}\BibitemShut {NoStop}%
\bibitem [{\citenamefont {Zhang}\ \emph {et~al.}(2009)\citenamefont {Zhang},
  \citenamefont {Liu}, \citenamefont {Qi}, \citenamefont {Dai}, \citenamefont
  {Fang},\ and\ \citenamefont {Zhang}}]{Zhang2009}%
  \BibitemOpen
  \bibfield  {author} {\bibinfo {author} {\bibfnamefont {Haijun}\ \bibnamefont
  {Zhang}}, \bibinfo {author} {\bibfnamefont {Chao-Xing}\ \bibnamefont {Liu}},
  \bibinfo {author} {\bibfnamefont {Xiao-Liang}\ \bibnamefont {Qi}}, \bibinfo
  {author} {\bibfnamefont {Xi}~\bibnamefont {Dai}}, \bibinfo {author}
  {\bibfnamefont {Zhong}\ \bibnamefont {Fang}}, \ and\ \bibinfo {author}
  {\bibfnamefont {Shou-Cheng}\ \bibnamefont {Zhang}},\ }\bibfield  {title}
  {\enquote {\bibinfo {title} {{Topological insulators in Bi2Se3, Bi2Te3 and
  Sb2Te3 with a single Dirac cone on the surface}},}\ }\href {\doibase
  10.1038/nphys1270} {\bibfield  {journal} {\bibinfo  {journal} {Nat. Phys.}\
  }\textbf {\bibinfo {volume} {5}},\ \bibinfo {pages} {438--442} (\bibinfo
  {year} {2009})}\BibitemShut {NoStop}%
\bibitem [{\citenamefont {Chen}\ \emph {et~al.}(2015)\citenamefont {Chen},
  \citenamefont {Jauregui}, \citenamefont {Tan}, \citenamefont {Manfra},
  \citenamefont {Klimeck}, \citenamefont {Chen},\ and\ \citenamefont
  {Kubis}}]{Chen2015}%
  \BibitemOpen
  \bibfield  {author} {\bibinfo {author} {\bibfnamefont {Fan~W.}\ \bibnamefont
  {Chen}}, \bibinfo {author} {\bibfnamefont {Luis~a.}\ \bibnamefont
  {Jauregui}}, \bibinfo {author} {\bibfnamefont {Yaohua}\ \bibnamefont {Tan}},
  \bibinfo {author} {\bibfnamefont {Michael}\ \bibnamefont {Manfra}}, \bibinfo
  {author} {\bibfnamefont {Gerhard}\ \bibnamefont {Klimeck}}, \bibinfo {author}
  {\bibfnamefont {Yong~P.}\ \bibnamefont {Chen}}, \ and\ \bibinfo {author}
  {\bibfnamefont {Tillmann}\ \bibnamefont {Kubis}},\ }\bibfield  {title}
  {\enquote {\bibinfo {title} {{In-surface confinement of topological insulator
  nanowire surface states}},}\ }\href {\doibase 10.1063/1.4931975} {\bibfield
  {journal} {\bibinfo  {journal} {Appl. Phys. Lett.}\ }\textbf {\bibinfo
  {volume} {107}},\ \bibinfo {pages} {121605} (\bibinfo {year}
  {2015})}\BibitemShut {NoStop}%
\bibitem [{\citenamefont {Dufouleur}\ \emph {et~al.}(2017)\citenamefont
  {Dufouleur}, \citenamefont {Veyrat}, \citenamefont {Dassonneville},
  \citenamefont {Xypakis}, \citenamefont {Bardarson}, \citenamefont {Nowka},
  \citenamefont {Hampel}, \citenamefont {Schumann}, \citenamefont {Eichler},
  \citenamefont {Schmidt}, \citenamefont {B{\"{u}}chner},\ and\ \citenamefont
  {Giraud}}]{Dufouleur2017}%
  \BibitemOpen
  \bibfield  {author} {\bibinfo {author} {\bibfnamefont {J.}~\bibnamefont
  {Dufouleur}}, \bibinfo {author} {\bibfnamefont {L.}~\bibnamefont {Veyrat}},
  \bibinfo {author} {\bibfnamefont {B.}~\bibnamefont {Dassonneville}}, \bibinfo
  {author} {\bibfnamefont {E.}~\bibnamefont {Xypakis}}, \bibinfo {author}
  {\bibfnamefont {J.~H.}\ \bibnamefont {Bardarson}}, \bibinfo {author}
  {\bibfnamefont {C.}~\bibnamefont {Nowka}}, \bibinfo {author} {\bibfnamefont
  {S.}~\bibnamefont {Hampel}}, \bibinfo {author} {\bibfnamefont
  {J.}~\bibnamefont {Schumann}}, \bibinfo {author} {\bibfnamefont
  {B.}~\bibnamefont {Eichler}}, \bibinfo {author} {\bibfnamefont {O.~G.}\
  \bibnamefont {Schmidt}}, \bibinfo {author} {\bibfnamefont {B.}~\bibnamefont
  {B{\"{u}}chner}}, \ and\ \bibinfo {author} {\bibfnamefont {R.}~\bibnamefont
  {Giraud}},\ }\bibfield  {title} {\enquote {\bibinfo {title} {{Weakly-coupled
  quasi-1D helical modes in disordered 3D topological insulator quantum
  wires}},}\ }\href {\doibase 10.1038/srep45276} {\bibfield  {journal}
  {\bibinfo  {journal} {Sci. Rep.}\ }\textbf {\bibinfo {volume} {7}},\ \bibinfo
  {pages} {45276} (\bibinfo {year} {2017})}\BibitemShut {NoStop}%
\bibitem [{\citenamefont {Cook}\ and\ \citenamefont {Franz}(2011)}]{Cook2011}%
  \BibitemOpen
  \bibfield  {author} {\bibinfo {author} {\bibfnamefont {A.}~\bibnamefont
  {Cook}}\ and\ \bibinfo {author} {\bibfnamefont {M.}~\bibnamefont {Franz}},\
  }\bibfield  {title} {\enquote {\bibinfo {title} {{Majorana fermions in a
  topological-insulator nanowire proximity-coupled to an s-wave
  superconductor}},}\ }\href {\doibase 10.1103/PhysRevB.84.201105} {\bibfield
  {journal} {\bibinfo  {journal} {Phys. Rev. B}\ }\textbf {\bibinfo {volume}
  {84}},\ \bibinfo {pages} {201105(R)} (\bibinfo {year} {2011})}\BibitemShut
  {NoStop}%
\bibitem [{\citenamefont {{de Juan}}\ \emph {et~al.}(2014)\citenamefont {{de
  Juan}}, \citenamefont {Ilan},\ and\ \citenamefont {Bardarson}}]{DeJuan2014}%
  \BibitemOpen
  \bibfield  {author} {\bibinfo {author} {\bibfnamefont {Fernando}\
  \bibnamefont {{de Juan}}}, \bibinfo {author} {\bibfnamefont {Roni}\
  \bibnamefont {Ilan}}, \ and\ \bibinfo {author} {\bibfnamefont {Jens~H.}\
  \bibnamefont {Bardarson}},\ }\bibfield  {title} {\enquote {\bibinfo {title}
  {{Robust transport signatures of topological superconductivity in topological
  insulator nanowires}},}\ }\href {\doibase 10.1103/PhysRevLett.113.107003}
  {\bibfield  {journal} {\bibinfo  {journal} {Phys. Rev. Lett.}\ }\textbf
  {\bibinfo {volume} {113}},\ \bibinfo {pages} {107003} (\bibinfo {year}
  {2014})}\BibitemShut {NoStop}%
\bibitem [{\citenamefont {de~Juan}\ \emph {et~al.}(2011)\citenamefont
  {de~Juan}, \citenamefont {Cortijo}, \citenamefont {Vozmediano},\ and\
  \citenamefont {Cano}}]{DeJuan2011}%
  \BibitemOpen
  \bibfield  {author} {\bibinfo {author} {\bibfnamefont {Fernando}\
  \bibnamefont {de~Juan}}, \bibinfo {author} {\bibfnamefont {Alberto}\
  \bibnamefont {Cortijo}}, \bibinfo {author} {\bibfnamefont {Mar{\'{i}}a
  A.~H.}\ \bibnamefont {Vozmediano}}, \ and\ \bibinfo {author} {\bibfnamefont
  {Andr{\'{e}}s}\ \bibnamefont {Cano}},\ }\bibfield  {title} {\enquote
  {\bibinfo {title} {{Aharonov-Bohm interferences from local deformations in
  graphene}},}\ }\href {\doibase 10.1038/nphys2034} {\bibfield  {journal}
  {\bibinfo  {journal} {Nat. Phys.}\ }\textbf {\bibinfo {volume} {7}},\
  \bibinfo {pages} {810} (\bibinfo {year} {2011})}\BibitemShut {NoStop}%
\bibitem [{\citenamefont {Levy}\ \emph {et~al.}(2010)\citenamefont {Levy},
  \citenamefont {Burke}, \citenamefont {Meaker}, \citenamefont {Panlasigui},
  \citenamefont {Zettl}, \citenamefont {Guinea}, \citenamefont {Neto},\ and\
  \citenamefont {Crommie}}]{Levy2010}%
  \BibitemOpen
  \bibfield  {author} {\bibinfo {author} {\bibfnamefont {N.}~\bibnamefont
  {Levy}}, \bibinfo {author} {\bibfnamefont {S.~A.}\ \bibnamefont {Burke}},
  \bibinfo {author} {\bibfnamefont {K.~L.}\ \bibnamefont {Meaker}}, \bibinfo
  {author} {\bibfnamefont {M.}~\bibnamefont {Panlasigui}}, \bibinfo {author}
  {\bibfnamefont {A.}~\bibnamefont {Zettl}}, \bibinfo {author} {\bibfnamefont
  {F.}~\bibnamefont {Guinea}}, \bibinfo {author} {\bibfnamefont {A.~H.~C.}\
  \bibnamefont {Neto}}, \ and\ \bibinfo {author} {\bibfnamefont {M.~F.}\
  \bibnamefont {Crommie}},\ }\bibfield  {title} {\enquote {\bibinfo {title}
  {{Strain-Induced Pseudo-Magnetic Fields Greater Than 300 Tesla in Graphene
  Nanobubbles}},}\ }\href {\doibase 10.1126/science.1191700} {\bibfield
  {journal} {\bibinfo  {journal} {Science}\ }\textbf {\bibinfo {volume}
  {329}},\ \bibinfo {pages} {544--547} (\bibinfo {year} {2010})}\BibitemShut
  {NoStop}%
\bibitem [{\citenamefont {Vozmediano}\ \emph {et~al.}(2010)\citenamefont
  {Vozmediano}, \citenamefont {Katsnelson},\ and\ \citenamefont
  {Guinea}}]{Vozmediano2010}%
  \BibitemOpen
  \bibfield  {author} {\bibinfo {author} {\bibfnamefont {M.A.H.}\ \bibnamefont
  {Vozmediano}}, \bibinfo {author} {\bibfnamefont {M.I.}\ \bibnamefont
  {Katsnelson}}, \ and\ \bibinfo {author} {\bibfnamefont {F.}~\bibnamefont
  {Guinea}},\ }\bibfield  {title} {\enquote {\bibinfo {title} {{Gauge fields in
  graphene}},}\ }\href {\doibase 10.1016/j.physrep.2010.07.003} {\bibfield
  {journal} {\bibinfo  {journal} {Phys. Rep.}\ }\textbf {\bibinfo {volume}
  {496}},\ \bibinfo {pages} {109--148} (\bibinfo {year} {2010})}\BibitemShut
  {NoStop}%
\bibitem [{\citenamefont {Amorim}\ \emph {et~al.}(2015)\citenamefont {Amorim},
  \citenamefont {Cortijo}, \citenamefont {{De Juan}}, \citenamefont {Grushin},
  \citenamefont {Guinea}, \citenamefont {Guti{\'{e}}rrez-Rubio}, \citenamefont
  {Ochoa}, \citenamefont {Parente}, \citenamefont {Rold{\'{a}}n}, \citenamefont
  {San-Jose}, \citenamefont {Schiefele}, \citenamefont {Sturla},\ and\
  \citenamefont {Vozmediano}}]{Amorim2016}%
  \BibitemOpen
  \bibfield  {author} {\bibinfo {author} {\bibfnamefont {B}~\bibnamefont
  {Amorim}}, \bibinfo {author} {\bibfnamefont {A}~\bibnamefont {Cortijo}},
  \bibinfo {author} {\bibfnamefont {F.}~\bibnamefont {{De Juan}}}, \bibinfo
  {author} {\bibfnamefont {A~G}\ \bibnamefont {Grushin}}, \bibinfo {author}
  {\bibfnamefont {F.}~\bibnamefont {Guinea}}, \bibinfo {author} {\bibfnamefont
  {A.}~\bibnamefont {Guti{\'{e}}rrez-Rubio}}, \bibinfo {author} {\bibfnamefont
  {H.}~\bibnamefont {Ochoa}}, \bibinfo {author} {\bibfnamefont
  {V.}~\bibnamefont {Parente}}, \bibinfo {author} {\bibfnamefont
  {R.}~\bibnamefont {Rold{\'{a}}n}}, \bibinfo {author} {\bibfnamefont
  {P.}~\bibnamefont {San-Jose}}, \bibinfo {author} {\bibfnamefont
  {J.}~\bibnamefont {Schiefele}}, \bibinfo {author} {\bibfnamefont
  {M.}~\bibnamefont {Sturla}}, \ and\ \bibinfo {author} {\bibfnamefont
  {M.~A.H.}\ \bibnamefont {Vozmediano}},\ }\bibfield  {title} {\enquote
  {\bibinfo {title} {{Novel effects of strains in graphene and other two
  dimensional materials}},}\ }\href {\doibase 10.1016/j.physrep.2015.12.006}
  {\bibfield  {journal} {\bibinfo  {journal} {Phys. Rep.}\ }\textbf {\bibinfo
  {volume} {617}},\ \bibinfo {pages} {1--54} (\bibinfo {year}
  {2015})}\BibitemShut {NoStop}%
\bibitem [{\citenamefont {Jeong}\ \emph {et~al.}(2011)\citenamefont {Jeong},
  \citenamefont {Shin},\ and\ \citenamefont {Lee}}]{Jeong2011}%
  \BibitemOpen
  \bibfield  {author} {\bibinfo {author} {\bibfnamefont {Jae~Seung}\
  \bibnamefont {Jeong}}, \bibinfo {author} {\bibfnamefont {Jeongkyu}\
  \bibnamefont {Shin}}, \ and\ \bibinfo {author} {\bibfnamefont {Hyun~Woo}\
  \bibnamefont {Lee}},\ }\bibfield  {title} {\enquote {\bibinfo {title}
  {{Curvature-induced spin-orbit coupling and spin relaxation in a chemically
  clean single-layer graphene}},}\ }\href {\doibase 10.1103/PhysRevB.84.195457}
  {\bibfield  {journal} {\bibinfo  {journal} {Phys. Rev. B}\ }\textbf {\bibinfo
  {volume} {84}},\ \bibinfo {pages} {195457} (\bibinfo {year}
  {2011})}\BibitemShut {NoStop}%
\bibitem [{\citenamefont {Yan}\ \emph {et~al.}(2013)\citenamefont {Yan},
  \citenamefont {He}, \citenamefont {Chu}, \citenamefont {Liu}, \citenamefont
  {Meng}, \citenamefont {Dou}, \citenamefont {Zhang}, \citenamefont {Liu},
  \citenamefont {Nie},\ and\ \citenamefont {He}}]{Yan2013}%
  \BibitemOpen
  \bibfield  {author} {\bibinfo {author} {\bibfnamefont {Wei}\ \bibnamefont
  {Yan}}, \bibinfo {author} {\bibfnamefont {Wen-Yu}\ \bibnamefont {He}},
  \bibinfo {author} {\bibfnamefont {Zhao-Dong}\ \bibnamefont {Chu}}, \bibinfo
  {author} {\bibfnamefont {Mengxi}\ \bibnamefont {Liu}}, \bibinfo {author}
  {\bibfnamefont {Lan}\ \bibnamefont {Meng}}, \bibinfo {author} {\bibfnamefont
  {Rui-Fen}\ \bibnamefont {Dou}}, \bibinfo {author} {\bibfnamefont {Yanfeng}\
  \bibnamefont {Zhang}}, \bibinfo {author} {\bibfnamefont {Zhongfan}\
  \bibnamefont {Liu}}, \bibinfo {author} {\bibfnamefont {Jia-Cai}\ \bibnamefont
  {Nie}}, \ and\ \bibinfo {author} {\bibfnamefont {Lin}\ \bibnamefont {He}},\
  }\bibfield  {title} {\enquote {\bibinfo {title} {{Strain and curvature
  induced evolution of electronic band structures in twisted graphene
  bilayer.}}}\ }\href {\doibase 10.1038/ncomms3159} {\bibfield  {journal}
  {\bibinfo  {journal} {Nat. Commun.}\ }\textbf {\bibinfo {volume} {4}},\
  \bibinfo {pages} {2159} (\bibinfo {year} {2013})}\BibitemShut {NoStop}%
\bibitem [{\citenamefont {Zwierzycki}(2014)}]{Zwierzycki2014}%
  \BibitemOpen
  \bibfield  {author} {\bibinfo {author} {\bibfnamefont {Maciej}\ \bibnamefont
  {Zwierzycki}},\ }\bibfield  {title} {\enquote {\bibinfo {title} {{Transport
  properties of rippled graphene}},}\ }\href
  {http://iopscience.iop.org/article/10.1088/0953-8984/26/13/135303/meta}
  {\bibfield  {journal} {\bibinfo  {journal} {J. Phys. Condens. Matter}\
  }\textbf {\bibinfo {volume} {26}},\ \bibinfo {pages} {135303} (\bibinfo
  {year} {2014})}\BibitemShut {NoStop}%
\bibitem [{\citenamefont {Burgos}\ \emph {et~al.}(2015)\citenamefont {Burgos},
  \citenamefont {Warnes}, \citenamefont {Lima},\ and\ \citenamefont
  {Lewenkopf}}]{Burgos2015}%
  \BibitemOpen
  \bibfield  {author} {\bibinfo {author} {\bibfnamefont {Rhonald}\ \bibnamefont
  {Burgos}}, \bibinfo {author} {\bibfnamefont {Jesus}\ \bibnamefont {Warnes}},
  \bibinfo {author} {\bibfnamefont {Leandro R.~F.}\ \bibnamefont {Lima}}, \
  and\ \bibinfo {author} {\bibfnamefont {Caio}\ \bibnamefont {Lewenkopf}},\
  }\bibfield  {title} {\enquote {\bibinfo {title} {{Effects of a random gauge
  field on the conductivity of graphene sheets with disordered ripples}},}\
  }\href {\doibase 10.1103/PhysRevB.91.115403} {\bibfield  {journal} {\bibinfo
  {journal} {Phys. Rev. B}\ }\textbf {\bibinfo {volume} {91}},\ \bibinfo
  {pages} {115403} (\bibinfo {year} {2015})}\BibitemShut {NoStop}%
\bibitem [{\citenamefont {Guan}\ and\ \citenamefont {Du}(2017)}]{Guan2017}%
  \BibitemOpen
  \bibfield  {author} {\bibinfo {author} {\bibfnamefont {Fen}\ \bibnamefont
  {Guan}}\ and\ \bibinfo {author} {\bibfnamefont {Xu}~\bibnamefont {Du}},\
  }\bibfield  {title} {\enquote {\bibinfo {title} {{Random Gauge Field
  Scattering in Monolayer Graphene}},}\ }\href {\doibase
  10.1021/acs.nanolett.7b03618} {\bibfield  {journal} {\bibinfo  {journal}
  {Nano Lett.}\ }\textbf {\bibinfo {volume} {17}},\ \bibinfo {pages}
  {7009--7014} (\bibinfo {year} {2017})}\BibitemShut {NoStop}%
\bibitem [{\citenamefont {Tang}\ and\ \citenamefont {Fu}(2014)}]{Tang2014a}%
  \BibitemOpen
  \bibfield  {author} {\bibinfo {author} {\bibfnamefont {Evelyn}\ \bibnamefont
  {Tang}}\ and\ \bibinfo {author} {\bibfnamefont {Liang}\ \bibnamefont {Fu}},\
  }\bibfield  {title} {\enquote {\bibinfo {title} {{Strain-induced partially
  flat band, helical snake states and interface superconductivity in
  topological crystalline insulators}},}\ }\href {\doibase 10.1038/nphys3109}
  {\bibfield  {journal} {\bibinfo  {journal} {Nat. Phys.}\ }\textbf {\bibinfo
  {volume} {10}},\ \bibinfo {pages} {964--969} (\bibinfo {year}
  {2014})}\BibitemShut {NoStop}%
\bibitem [{\citenamefont {Cortijo}\ \emph {et~al.}(2015)\citenamefont
  {Cortijo}, \citenamefont {Ferreir{\'{o}}s}, \citenamefont {Landsteiner},\
  and\ \citenamefont {Vozmediano}}]{Cortijo2015}%
  \BibitemOpen
  \bibfield  {author} {\bibinfo {author} {\bibfnamefont {Alberto}\ \bibnamefont
  {Cortijo}}, \bibinfo {author} {\bibfnamefont {Yago}\ \bibnamefont
  {Ferreir{\'{o}}s}}, \bibinfo {author} {\bibfnamefont {Karl}\ \bibnamefont
  {Landsteiner}}, \ and\ \bibinfo {author} {\bibfnamefont {Mar{\'{i}}a A.~H.}\
  \bibnamefont {Vozmediano}},\ }\bibfield  {title} {\enquote {\bibinfo {title}
  {{Elastic gauge fields in Weyl semimetals}},}\ }\href {\doibase
  10.1103/PhysRevLett.115.177202} {\bibfield  {journal} {\bibinfo  {journal}
  {Phys. Rev. Lett.}\ }\textbf {\bibinfo {volume} {115}},\ \bibinfo {pages}
  {177202} (\bibinfo {year} {2015})}\BibitemShut {NoStop}%
\bibitem [{\citenamefont {Takane}\ and\ \citenamefont {Imura}(2013)}]{Takane}%
  \BibitemOpen
  \bibfield  {author} {\bibinfo {author} {\bibfnamefont {Yositake}\
  \bibnamefont {Takane}}\ and\ \bibinfo {author} {\bibfnamefont {Ken-ichiro}\
  \bibnamefont {Imura}},\ }\bibfield  {title} {\enquote {\bibinfo {title}
  {{Unified Description of Dirac Electrons on a Curved Surface of Topological
  Insulators}},}\ }\href {\doibase 10.7566/JPSJ.82.074712} {\bibfield
  {journal} {\bibinfo  {journal} {J. Phys. Soc. Japan}\ }\textbf {\bibinfo
  {volume} {82}},\ \bibinfo {pages} {074712} (\bibinfo {year}
  {2013})}\BibitemShut {NoStop}%
\bibitem [{\citenamefont {Utiyama}(1956)}]{Utiyama1956}%
  \BibitemOpen
  \bibfield  {author} {\bibinfo {author} {\bibfnamefont {Ryoyu}\ \bibnamefont
  {Utiyama}},\ }\bibfield  {title} {\enquote {\bibinfo {title} {{Invariant
  theoretical interpretation of interaction}},}\ }\href {\doibase
  10.1103/PhysRev.101.1597} {\bibfield  {journal} {\bibinfo  {journal} {Phys.
  Rev.}\ }\textbf {\bibinfo {volume} {101}},\ \bibinfo {pages} {1597--1607}
  (\bibinfo {year} {1956})}\BibitemShut {NoStop}%
\bibitem [{\citenamefont {Schnyder}\ \emph {et~al.}(2009)\citenamefont
  {Schnyder}, \citenamefont {Ryu}, \citenamefont {Furusaki},\ and\
  \citenamefont {Ludwig}}]{Schnyder2009a}%
  \BibitemOpen
  \bibfield  {author} {\bibinfo {author} {\bibfnamefont {Andreas~P.}\
  \bibnamefont {Schnyder}}, \bibinfo {author} {\bibfnamefont {Shinsei}\
  \bibnamefont {Ryu}}, \bibinfo {author} {\bibfnamefont {Akira}\ \bibnamefont
  {Furusaki}}, \ and\ \bibinfo {author} {\bibfnamefont {Andreas~W.W.}\
  \bibnamefont {Ludwig}},\ }\bibfield  {title} {\enquote {\bibinfo {title}
  {{Classification of topological insulators and superconductors}},}\ }\href
  {\doibase 10.1063/1.3149481} {\bibfield  {journal} {\bibinfo  {journal} {AIP
  Conf. Proc.}\ }\textbf {\bibinfo {volume} {1134}},\ \bibinfo {pages} {10--21}
  (\bibinfo {year} {2009})}\BibitemShut {NoStop}%
\bibitem [{\citenamefont {Chiu}\ \emph {et~al.}(2016)\citenamefont {Chiu},
  \citenamefont {Teo}, \citenamefont {Schnyder},\ and\ \citenamefont
  {Ryu}}]{Chiu2016}%
  \BibitemOpen
  \bibfield  {author} {\bibinfo {author} {\bibfnamefont {Ching-Kai}\
  \bibnamefont {Chiu}}, \bibinfo {author} {\bibfnamefont {Jeffrey C.~Y.}\
  \bibnamefont {Teo}}, \bibinfo {author} {\bibfnamefont {Andreas~P.}\
  \bibnamefont {Schnyder}}, \ and\ \bibinfo {author} {\bibfnamefont {Shinsei}\
  \bibnamefont {Ryu}},\ }\bibfield  {title} {\enquote {\bibinfo {title}
  {{Classification of topological quantum matter with symmetries}},}\ }\href
  {\doibase 10.1103/RevModPhys.88.035005} {\bibfield  {journal} {\bibinfo
  {journal} {Rev. Mod. Phys.}\ }\textbf {\bibinfo {volume} {88}},\ \bibinfo
  {pages} {035005} (\bibinfo {year} {2016})}\BibitemShut {NoStop}%
\bibitem [{\citenamefont {Jauregui}\ \emph {et~al.}(2018)\citenamefont
  {Jauregui}, \citenamefont {Kayyalha}, \citenamefont {Kazakov}, \citenamefont
  {Miotkowski}, \citenamefont {Rokhinson},\ and\ \citenamefont
  {Chen}}]{Jauregui2018}%
  \BibitemOpen
  \bibfield  {author} {\bibinfo {author} {\bibfnamefont {Luis~A.}\ \bibnamefont
  {Jauregui}}, \bibinfo {author} {\bibfnamefont {Morteza}\ \bibnamefont
  {Kayyalha}}, \bibinfo {author} {\bibfnamefont {Aleksandr}\ \bibnamefont
  {Kazakov}}, \bibinfo {author} {\bibfnamefont {Ireneusz}\ \bibnamefont
  {Miotkowski}}, \bibinfo {author} {\bibfnamefont {Leonid~P.}\ \bibnamefont
  {Rokhinson}}, \ and\ \bibinfo {author} {\bibfnamefont {Yong~P.}\ \bibnamefont
  {Chen}},\ }\bibfield  {title} {\enquote {\bibinfo {title} {{Gate-tunable
  supercurrent and multiple Andreev reflections in a superconductor-topological
  insulator nanoribbon-superconductor hybrid device}},}\ }\href {\doibase
  10.1063/1.5008746} {\bibfield  {journal} {\bibinfo  {journal} {Appl. Phys.
  Lett.}\ }\textbf {\bibinfo {volume} {112}},\ \bibinfo {pages} {093105}
  (\bibinfo {year} {2018})}\BibitemShut {NoStop}%
\bibitem [{\citenamefont {Kayyalha}\ \emph {et~al.}(2019)\citenamefont
  {Kayyalha}, \citenamefont {Kargarian}, \citenamefont {Kazakov}, \citenamefont
  {Miotkowski}, \citenamefont {Galitski}, \citenamefont {Yakovenko},
  \citenamefont {Rokhinson},\ and\ \citenamefont {Chen}}]{Kayyalha:2019gk}%
  \BibitemOpen
  \bibfield  {author} {\bibinfo {author} {\bibfnamefont {Morteza}\ \bibnamefont
  {Kayyalha}}, \bibinfo {author} {\bibfnamefont {Mehdi}\ \bibnamefont
  {Kargarian}}, \bibinfo {author} {\bibfnamefont {Aleksandr}\ \bibnamefont
  {Kazakov}}, \bibinfo {author} {\bibfnamefont {Ireneusz}\ \bibnamefont
  {Miotkowski}}, \bibinfo {author} {\bibfnamefont {Victor~M.}\ \bibnamefont
  {Galitski}}, \bibinfo {author} {\bibfnamefont {Victor~M.}\ \bibnamefont
  {Yakovenko}}, \bibinfo {author} {\bibfnamefont {Leonid~P.}\ \bibnamefont
  {Rokhinson}}, \ and\ \bibinfo {author} {\bibfnamefont {Yong~P.}\ \bibnamefont
  {Chen}},\ }\bibfield  {title} {\enquote {\bibinfo {title} {{Anomalous
  Low-Temperature Enhancement of Supercurrent in Topological-Insulator
  Nanoribbon Josephson Junctions: Evidence for Low-Energy Andreev Bound
  States}},}\ }\href {\doibase 10.1103/PhysRevLett.122.047003} {\bibfield
  {journal} {\bibinfo  {journal} {Phys. Rev. Lett.}\ }\textbf {\bibinfo
  {volume} {122}},\ \bibinfo {pages} {047003} (\bibinfo {year}
  {2019})}\BibitemShut {NoStop}%
\bibitem [{\citenamefont {Cai}\ \emph {et~al.}(2018)\citenamefont {Cai},
  \citenamefont {Guo}, \citenamefont {Sidorov}, \citenamefont {Zhou},
  \citenamefont {Wang}, \citenamefont {Lin}, \citenamefont {Li}, \citenamefont
  {Li}, \citenamefont {Yang}, \citenamefont {Li}, \citenamefont {Wu},
  \citenamefont {Hu}, \citenamefont {Kushwaha}, \citenamefont {Cava},\ and\
  \citenamefont {Sun}}]{Cai:2018dla}%
  \BibitemOpen
  \bibfield  {author} {\bibinfo {author} {\bibfnamefont {Shu}\ \bibnamefont
  {Cai}}, \bibinfo {author} {\bibfnamefont {Jing}\ \bibnamefont {Guo}},
  \bibinfo {author} {\bibfnamefont {Vladimir~A}\ \bibnamefont {Sidorov}},
  \bibinfo {author} {\bibfnamefont {Yazhou}\ \bibnamefont {Zhou}}, \bibinfo
  {author} {\bibfnamefont {Honghong}\ \bibnamefont {Wang}}, \bibinfo {author}
  {\bibfnamefont {Gongchang}\ \bibnamefont {Lin}}, \bibinfo {author}
  {\bibfnamefont {Xiaodong}\ \bibnamefont {Li}}, \bibinfo {author}
  {\bibfnamefont {Yanchuan}\ \bibnamefont {Li}}, \bibinfo {author}
  {\bibfnamefont {Ke}~\bibnamefont {Yang}}, \bibinfo {author} {\bibfnamefont
  {Aiguo}\ \bibnamefont {Li}}, \bibinfo {author} {\bibfnamefont
  {Qi}~\bibnamefont {Wu}}, \bibinfo {author} {\bibfnamefont {Jiangping}\
  \bibnamefont {Hu}}, \bibinfo {author} {\bibfnamefont {Satya~K}\ \bibnamefont
  {Kushwaha}}, \bibinfo {author} {\bibfnamefont {Robert~J}\ \bibnamefont
  {Cava}}, \ and\ \bibinfo {author} {\bibfnamefont {Liling}\ \bibnamefont
  {Sun}},\ }\bibfield  {title} {\enquote {\bibinfo {title} {{Independence of
  topological surface state and bulk conductance in three-dimensional
  topological insulators}},}\ }\href {\doibase 10.1038/s41535-018-0134-z}
  {\bibfield  {journal} {\bibinfo  {journal} {npj Quantum Materials}\ }\textbf
  {\bibinfo {volume} {3}},\ \bibinfo {pages} {62} (\bibinfo {year}
  {2018})}\BibitemShut {NoStop}%
\bibitem [{\citenamefont {Xypakis}\ and\ \citenamefont
  {Bardarson}(2017)}]{Xypakis2017}%
  \BibitemOpen
  \bibfield  {author} {\bibinfo {author} {\bibfnamefont {Emmanouil}\
  \bibnamefont {Xypakis}}\ and\ \bibinfo {author} {\bibfnamefont {Jens~H.}\
  \bibnamefont {Bardarson}},\ }\bibfield  {title} {\enquote {\bibinfo {title}
  {{Conductance fluctuations and disorder induced $\nu$=0 quantum Hall plateau
  in topological insulator nanowires}},}\ }\href {\doibase
  10.1103/PhysRevB.95.035415} {\bibfield  {journal} {\bibinfo  {journal} {Phys.
  Rev. B}\ }\textbf {\bibinfo {volume} {95}},\ \bibinfo {pages} {035415}
  (\bibinfo {year} {2017})}\BibitemShut {NoStop}%
\bibitem [{\citenamefont {Beenakker}(1997)}]{Beenakker1997}%
  \BibitemOpen
  \bibfield  {author} {\bibinfo {author} {\bibfnamefont {C.~W.~J.}\
  \bibnamefont {Beenakker}},\ }\bibfield  {title} {\enquote {\bibinfo {title}
  {{Random-matrix theory of quantum transport}},}\ }\href {\doibase
  10.1103/RevModPhys.69.731} {\bibfield  {journal} {\bibinfo  {journal} {Rev.
  Mod. Phys.}\ }\textbf {\bibinfo {volume} {69}},\ \bibinfo {pages} {731--808}
  (\bibinfo {year} {1997})}\BibitemShut {NoStop}%
\bibitem [{\citenamefont {Mello}\ \emph {et~al.}(1988)\citenamefont {Mello},
  \citenamefont {Pereyra},\ and\ \citenamefont {Kumar}}]{Mello1988}%
  \BibitemOpen
  \bibfield  {author} {\bibinfo {author} {\bibfnamefont {P.A}\ \bibnamefont
  {Mello}}, \bibinfo {author} {\bibfnamefont {P.}~\bibnamefont {Pereyra}}, \
  and\ \bibinfo {author} {\bibfnamefont {N.}~\bibnamefont {Kumar}},\ }\bibfield
   {title} {\enquote {\bibinfo {title} {{Macroscopic approach to multichannel
  disordered conductors}},}\ }\href {\doibase 10.1016/0003-4916(88)90169-8}
  {\bibfield  {journal} {\bibinfo  {journal} {Ann. Phys. (N. Y).}\ }\textbf
  {\bibinfo {volume} {181}},\ \bibinfo {pages} {290--317} (\bibinfo {year}
  {1988})}\BibitemShut {NoStop}%
\bibitem [{\citenamefont {Tworzyd{\l}o}\ \emph {et~al.}(2006)\citenamefont
  {Tworzyd{\l}o}, \citenamefont {Trauzettel}, \citenamefont {Titov},
  \citenamefont {Rycerz},\ and\ \citenamefont {Beenakker}}]{Tworzydo2006}%
  \BibitemOpen
  \bibfield  {author} {\bibinfo {author} {\bibfnamefont {J.}~\bibnamefont
  {Tworzyd{\l}o}}, \bibinfo {author} {\bibfnamefont {B.}~\bibnamefont
  {Trauzettel}}, \bibinfo {author} {\bibfnamefont {M.}~\bibnamefont {Titov}},
  \bibinfo {author} {\bibfnamefont {A.}~\bibnamefont {Rycerz}}, \ and\ \bibinfo
  {author} {\bibfnamefont {C.~W~J}\ \bibnamefont {Beenakker}},\ }\bibfield
  {title} {\enquote {\bibinfo {title} {{Sub-Poissonian Shot Noise in
  Graphene}},}\ }\href {\doibase 10.1103/PhysRevLett.96.246802} {\bibfield
  {journal} {\bibinfo  {journal} {Phys. Rev. Lett.}\ }\textbf {\bibinfo
  {volume} {96}},\ \bibinfo {pages} {246802} (\bibinfo {year}
  {2006})}\BibitemShut {NoStop}%
\bibitem [{\citenamefont {Fulga}\ \emph
  {et~al.}(2011{\natexlab{a}})\citenamefont {Fulga}, \citenamefont {Hassler},
  \citenamefont {Akhmerov},\ and\ \citenamefont {Beenakker}}]{Fulga2011a}%
  \BibitemOpen
  \bibfield  {author} {\bibinfo {author} {\bibfnamefont {I.~C.}\ \bibnamefont
  {Fulga}}, \bibinfo {author} {\bibfnamefont {F.}~\bibnamefont {Hassler}},
  \bibinfo {author} {\bibfnamefont {A.~R.}\ \bibnamefont {Akhmerov}}, \ and\
  \bibinfo {author} {\bibfnamefont {C.~W.~J.}\ \bibnamefont {Beenakker}},\
  }\bibfield  {title} {\enquote {\bibinfo {title} {{Scattering formula for the
  topological quantum number of a disordered multimode wire}},}\ }\href
  {\doibase 10.1103/PhysRevB.83.155429} {\bibfield  {journal} {\bibinfo
  {journal} {Phys. Rev. B}\ }\textbf {\bibinfo {volume} {83}},\ \bibinfo
  {pages} {155429} (\bibinfo {year} {2011}{\natexlab{a}})}\BibitemShut
  {NoStop}%
\bibitem [{\citenamefont {Fulga}\ \emph
  {et~al.}(2011{\natexlab{b}})\citenamefont {Fulga}, \citenamefont {Hassler},
  \citenamefont {Akhmerov},\ and\ \citenamefont {Beenakker}}]{Fulga2011}%
  \BibitemOpen
  \bibfield  {author} {\bibinfo {author} {\bibfnamefont {I.~C.}\ \bibnamefont
  {Fulga}}, \bibinfo {author} {\bibfnamefont {F.}~\bibnamefont {Hassler}},
  \bibinfo {author} {\bibfnamefont {A.~R.}\ \bibnamefont {Akhmerov}}, \ and\
  \bibinfo {author} {\bibfnamefont {C.~W~J}\ \bibnamefont {Beenakker}},\
  }\bibfield  {title} {\enquote {\bibinfo {title} {{Topological quantum number
  and critical exponent from conductance fluctuations at the quantum Hall
  plateau transition}},}\ }\href {\doibase 10.1103/PhysRevB.84.245447}
  {\bibfield  {journal} {\bibinfo  {journal} {Phys. Rev. B}\ }\textbf {\bibinfo
  {volume} {84}},\ \bibinfo {pages} {245447} (\bibinfo {year}
  {2011}{\natexlab{b}})}\BibitemShut {NoStop}%
\bibitem [{\citenamefont {Altland}\ and\ \citenamefont
  {Zirnbauer}(1997)}]{Altland1997}%
  \BibitemOpen
  \bibfield  {author} {\bibinfo {author} {\bibfnamefont {Alexander}\
  \bibnamefont {Altland}}\ and\ \bibinfo {author} {\bibfnamefont {Martin~R}\
  \bibnamefont {Zirnbauer}},\ }\bibfield  {title} {\enquote {\bibinfo {title}
  {{Nonstandard symmetry classes in mesoscopic normal-superconducting hybrid
  structures}},}\ }\href {\doibase 10.1103/PhysRevB.55.1142} {\bibfield
  {journal} {\bibinfo  {journal} {Phys. Rev. B}\ }\textbf {\bibinfo {volume}
  {55}},\ \bibinfo {pages} {1142--1161} (\bibinfo {year} {1997})}\BibitemShut
  {NoStop}%
\bibitem [{\citenamefont {Sacksteder}\ and\ \citenamefont
  {Wu}(2016)}]{Sacksteder2016}%
  \BibitemOpen
  \bibfield  {author} {\bibinfo {author} {\bibfnamefont {Vincent~E.}\
  \bibnamefont {Sacksteder}}\ and\ \bibinfo {author} {\bibfnamefont
  {Quansheng}\ \bibnamefont {Wu}},\ }\bibfield  {title} {\enquote {\bibinfo
  {title} {{Quantum interference effects in topological nanowires in a
  longitudinal magnetic field}},}\ }\href {\doibase 10.1103/PhysRevB.94.205424}
  {\bibfield  {journal} {\bibinfo  {journal} {Phys. Rev. B}\ }\textbf {\bibinfo
  {volume} {94}},\ \bibinfo {pages} {205424} (\bibinfo {year}
  {2016})}\BibitemShut {NoStop}%
\bibitem [{\citenamefont {Cho}\ \emph {et~al.}(2015)\citenamefont {Cho},
  \citenamefont {Dellabetta}, \citenamefont {Zhong}, \citenamefont
  {Schneeloch}, \citenamefont {Liu}, \citenamefont {Gu}, \citenamefont
  {Gilbert},\ and\ \citenamefont {Mason}}]{Cho2015}%
  \BibitemOpen
  \bibfield  {author} {\bibinfo {author} {\bibfnamefont {Sungjae}\ \bibnamefont
  {Cho}}, \bibinfo {author} {\bibfnamefont {Brian}\ \bibnamefont {Dellabetta}},
  \bibinfo {author} {\bibfnamefont {Ruidan}\ \bibnamefont {Zhong}}, \bibinfo
  {author} {\bibfnamefont {John}\ \bibnamefont {Schneeloch}}, \bibinfo {author}
  {\bibfnamefont {Tiansheng}\ \bibnamefont {Liu}}, \bibinfo {author}
  {\bibfnamefont {Genda}\ \bibnamefont {Gu}}, \bibinfo {author} {\bibfnamefont
  {Matthew~J.}\ \bibnamefont {Gilbert}}, \ and\ \bibinfo {author}
  {\bibfnamefont {Nadya}\ \bibnamefont {Mason}},\ }\bibfield  {title} {\enquote
  {\bibinfo {title} {{Aharonov-Bohm oscillations in a quasi-ballistic
  three-dimensional topological insulator nanowire}},}\ }\href {\doibase
  10.1038/ncomms8634} {\bibfield  {journal} {\bibinfo  {journal} {Nat.
  Commun.}\ }\textbf {\bibinfo {volume} {6}},\ \bibinfo {pages} {7634}
  (\bibinfo {year} {2015})}\BibitemShut {NoStop}%
\bibitem [{\citenamefont {Jauregui}\ \emph {et~al.}(2016)\citenamefont
  {Jauregui}, \citenamefont {Pettes}, \citenamefont {Rokhinson}, \citenamefont
  {Shi},\ and\ \citenamefont {Chen}}]{Jauregui2016a}%
  \BibitemOpen
  \bibfield  {author} {\bibinfo {author} {\bibfnamefont {Luis~A}\ \bibnamefont
  {Jauregui}}, \bibinfo {author} {\bibfnamefont {Michael~T}\ \bibnamefont
  {Pettes}}, \bibinfo {author} {\bibfnamefont {Leonid~P}\ \bibnamefont
  {Rokhinson}}, \bibinfo {author} {\bibfnamefont {Li}~\bibnamefont {Shi}}, \
  and\ \bibinfo {author} {\bibfnamefont {Yong~P}\ \bibnamefont {Chen}},\
  }\bibfield  {title} {\enquote {\bibinfo {title} {Magnetic field-induced
  helical mode and topological transitions in a topological insulator
  nanoribbon},}\ }\href {\doibase 10.1038/nnano.2015.293} {\bibfield  {journal}
  {\bibinfo  {journal} {Nat. Nanotechnol.}\ }\textbf {\bibinfo {volume} {11}},\
  \bibinfo {pages} {345--351} (\bibinfo {year} {2016})}\BibitemShut {NoStop}%
\bibitem [{\citenamefont {Hong}\ \emph {et~al.}(2014)\citenamefont {Hong},
  \citenamefont {Zhang}, \citenamefont {Cha}, \citenamefont {Qi},\ and\
  \citenamefont {Cui}}]{Hong2014}%
  \BibitemOpen
  \bibfield  {author} {\bibinfo {author} {\bibfnamefont {Seung~Sae}\
  \bibnamefont {Hong}}, \bibinfo {author} {\bibfnamefont {Yi}~\bibnamefont
  {Zhang}}, \bibinfo {author} {\bibfnamefont {Judy~J.}\ \bibnamefont {Cha}},
  \bibinfo {author} {\bibfnamefont {Xiao~Liang}\ \bibnamefont {Qi}}, \ and\
  \bibinfo {author} {\bibfnamefont {Yi}~\bibnamefont {Cui}},\ }\bibfield
  {title} {\enquote {\bibinfo {title} {{One-dimensional helical transport in
  topological insulator nanowire interferometers}},}\ }\href {\doibase
  10.1021/nl500822g} {\bibfield  {journal} {\bibinfo  {journal} {Nano Lett.}\
  }\textbf {\bibinfo {volume} {14}},\ \bibinfo {pages} {2815--2821} (\bibinfo
  {year} {2014})}\BibitemShut {NoStop}%
\bibitem [{\citenamefont {Ziegler}\ \emph {et~al.}(2018)\citenamefont
  {Ziegler}, \citenamefont {Kozlovsky}, \citenamefont {Gorini}, \citenamefont
  {Liu}, \citenamefont {Weish{\"{a}}upl}, \citenamefont {Maier}, \citenamefont
  {Fischer}, \citenamefont {Kozlov}, \citenamefont {Kvon}, \citenamefont
  {Mikhailov}, \citenamefont {Dvoretsky}, \citenamefont {Richter},\ and\
  \citenamefont {Weiss}}]{Ziegler2018}%
  \BibitemOpen
  \bibfield  {author} {\bibinfo {author} {\bibfnamefont {J.}~\bibnamefont
  {Ziegler}}, \bibinfo {author} {\bibfnamefont {R.}~\bibnamefont {Kozlovsky}},
  \bibinfo {author} {\bibfnamefont {C.}~\bibnamefont {Gorini}}, \bibinfo
  {author} {\bibfnamefont {M.~H.}\ \bibnamefont {Liu}}, \bibinfo {author}
  {\bibfnamefont {S.}~\bibnamefont {Weish{\"{a}}upl}}, \bibinfo {author}
  {\bibfnamefont {H.}~\bibnamefont {Maier}}, \bibinfo {author} {\bibfnamefont
  {R.}~\bibnamefont {Fischer}}, \bibinfo {author} {\bibfnamefont {D.~A.}\
  \bibnamefont {Kozlov}}, \bibinfo {author} {\bibfnamefont {Z.~D.}\
  \bibnamefont {Kvon}}, \bibinfo {author} {\bibfnamefont {N.}~\bibnamefont
  {Mikhailov}}, \bibinfo {author} {\bibfnamefont {S.~A.}\ \bibnamefont
  {Dvoretsky}}, \bibinfo {author} {\bibfnamefont {K.}~\bibnamefont {Richter}},
  \ and\ \bibinfo {author} {\bibfnamefont {D.}~\bibnamefont {Weiss}},\
  }\bibfield  {title} {\enquote {\bibinfo {title} {{Probing spin helical
  surface states in topological HgTe nanowires}},}\ }\href {\doibase
  10.1103/PhysRevB.97.035157} {\bibfield  {journal} {\bibinfo  {journal} {Phys.
  Rev. B}\ }\textbf {\bibinfo {volume} {97}},\ \bibinfo {pages} {035157}
  (\bibinfo {year} {2018})}\BibitemShut {NoStop}%
\bibitem [{\citenamefont {Moors}\ \emph {et~al.}(2018)\citenamefont {Moors},
  \citenamefont {Sch{\"{u}}ffelgen}, \citenamefont {Rosenbach}, \citenamefont
  {Schmitt}, \citenamefont {Sch{\"{a}}pers},\ and\ \citenamefont
  {Schmidt}}]{Moors2018}%
  \BibitemOpen
  \bibfield  {author} {\bibinfo {author} {\bibfnamefont {Kristof}\ \bibnamefont
  {Moors}}, \bibinfo {author} {\bibfnamefont {Peter}\ \bibnamefont
  {Sch{\"{u}}ffelgen}}, \bibinfo {author} {\bibfnamefont {Daniel}\ \bibnamefont
  {Rosenbach}}, \bibinfo {author} {\bibfnamefont {Tobias}\ \bibnamefont
  {Schmitt}}, \bibinfo {author} {\bibfnamefont {Thomas}\ \bibnamefont
  {Sch{\"{a}}pers}}, \ and\ \bibinfo {author} {\bibfnamefont {Thomas~L.}\
  \bibnamefont {Schmidt}},\ }\bibfield  {title} {\enquote {\bibinfo {title}
  {{Magnetotransport signatures of 3D topological insulator nanowire
  structures}},}\ }\href {\doibase 10.1103/PhysRevB.97.245429} {\bibfield
  {journal} {\bibinfo  {journal} {Phys. Rev. B}\ }\textbf {\bibinfo {volume}
  {97}},\ \bibinfo {pages} {245429} (\bibinfo {year} {2018})}\BibitemShut
  {NoStop}%
\bibitem [{\citenamefont {Balents}\ and\ \citenamefont
  {Fisher}(1997)}]{balents1997delocalization}%
  \BibitemOpen
  \bibfield  {author} {\bibinfo {author} {\bibfnamefont {Leon}\ \bibnamefont
  {Balents}}\ and\ \bibinfo {author} {\bibfnamefont {Matthew P.~A.}\
  \bibnamefont {Fisher}},\ }\bibfield  {title} {\enquote {\bibinfo {title}
  {Delocalization transition via supersymmetry in one dimension},}\ }\href
  {\doibase 10.1103/PhysRevB.56.12970} {\bibfield  {journal} {\bibinfo
  {journal} {Phys. Rev. B}\ }\textbf {\bibinfo {volume} {56}},\ \bibinfo
  {pages} {12970} (\bibinfo {year} {1997})}\BibitemShut {NoStop}%
\bibitem [{\citenamefont {Brouwer}\ \emph {et~al.}(2000)\citenamefont
  {Brouwer}, \citenamefont {Furusaki}, \citenamefont {Gruzberg},\ and\
  \citenamefont {Mudry}}]{brouwer2000localization}%
  \BibitemOpen
  \bibfield  {author} {\bibinfo {author} {\bibfnamefont {P.~W.}\ \bibnamefont
  {Brouwer}}, \bibinfo {author} {\bibfnamefont {Akira}\ \bibnamefont
  {Furusaki}}, \bibinfo {author} {\bibfnamefont {I.~A.}\ \bibnamefont
  {Gruzberg}}, \ and\ \bibinfo {author} {\bibfnamefont {C}~\bibnamefont
  {Mudry}},\ }\bibfield  {title} {\enquote {\bibinfo {title} {Localization and
  delocalization in dirty superconducting wires},}\ }\href {\doibase
  10.1103/PhysRevLett.85.1064} {\bibfield  {journal} {\bibinfo  {journal}
  {Phys. Rev. Lett.}\ }\textbf {\bibinfo {volume} {85}},\ \bibinfo {pages}
  {1064} (\bibinfo {year} {2000})}\BibitemShut {NoStop}%
\bibitem [{\citenamefont {Motrunich}\ \emph {et~al.}(2001)\citenamefont
  {Motrunich}, \citenamefont {Damle},\ and\ \citenamefont
  {Huse}}]{motrunich2001griffiths}%
  \BibitemOpen
  \bibfield  {author} {\bibinfo {author} {\bibfnamefont {Olexei}\ \bibnamefont
  {Motrunich}}, \bibinfo {author} {\bibfnamefont {Kedar}\ \bibnamefont
  {Damle}}, \ and\ \bibinfo {author} {\bibfnamefont {David~A}\ \bibnamefont
  {Huse}},\ }\bibfield  {title} {\enquote {\bibinfo {title} {Griffiths effects
  and quantum critical points in dirty superconductors without spin-rotation
  invariance: One-dimensional examples},}\ }\href {\doibase
  10.1103/PhysRevB.63.224204} {\bibfield  {journal} {\bibinfo  {journal} {Phys.
  Rev. B}\ }\textbf {\bibinfo {volume} {63}},\ \bibinfo {pages} {224204}
  (\bibinfo {year} {2001})}\BibitemShut {NoStop}%
\bibitem [{\citenamefont {Gruzberg}\ \emph {et~al.}(2005)\citenamefont
  {Gruzberg}, \citenamefont {Read},\ and\ \citenamefont
  {Vishveshwara}}]{gruzberg2005localization}%
  \BibitemOpen
  \bibfield  {author} {\bibinfo {author} {\bibfnamefont {Ilya~A}\ \bibnamefont
  {Gruzberg}}, \bibinfo {author} {\bibfnamefont {N}~\bibnamefont {Read}}, \
  and\ \bibinfo {author} {\bibfnamefont {Smitha}\ \bibnamefont
  {Vishveshwara}},\ }\bibfield  {title} {\enquote {\bibinfo {title}
  {Localization in disordered superconducting wires with broken spin-rotation
  symmetry},}\ }\href {\doibase 10.1103/PhysRevB.71.245124} {\bibfield
  {journal} {\bibinfo  {journal} {Phys. Rev. B}\ }\textbf {\bibinfo {volume}
  {71}},\ \bibinfo {pages} {245124} (\bibinfo {year} {2005})}\BibitemShut
  {NoStop}%
\bibitem [{\citenamefont {Evers}\ and\ \citenamefont
  {Mirlin}(2008)}]{Evers2008}%
  \BibitemOpen
  \bibfield  {author} {\bibinfo {author} {\bibfnamefont {Ferdinand}\
  \bibnamefont {Evers}}\ and\ \bibinfo {author} {\bibfnamefont {Alexander~D.}\
  \bibnamefont {Mirlin}},\ }\bibfield  {title} {\enquote {\bibinfo {title}
  {{Anderson transitions}},}\ }\href {\doibase 10.1103/RevModPhys.80.1355}
  {\bibfield  {journal} {\bibinfo  {journal} {Rev. Mod. Phys.}\ }\textbf
  {\bibinfo {volume} {80}},\ \bibinfo {pages} {1355--1417} (\bibinfo {year}
  {2008})}\BibitemShut {NoStop}%
\bibitem [{\citenamefont {Altland}\ \emph {et~al.}(2015)\citenamefont
  {Altland}, \citenamefont {Bagrets},\ and\ \citenamefont
  {Kamenev}}]{altland2015topology}%
  \BibitemOpen
  \bibfield  {author} {\bibinfo {author} {\bibfnamefont {Alexander}\
  \bibnamefont {Altland}}, \bibinfo {author} {\bibfnamefont {Dmitry}\
  \bibnamefont {Bagrets}}, \ and\ \bibinfo {author} {\bibfnamefont {Alex}\
  \bibnamefont {Kamenev}},\ }\bibfield  {title} {\enquote {\bibinfo {title}
  {Topology versus anderson localization: Nonperturbative solutions in one
  dimension},}\ }\href {\doibase 10.1103/PhysRevB.91.085429} {\bibfield
  {journal} {\bibinfo  {journal} {Phys. Rev. B}\ }\textbf {\bibinfo {volume}
  {91}},\ \bibinfo {pages} {085429} (\bibinfo {year} {2015})}\BibitemShut
  {NoStop}%
\end{thebibliography}%

\end{document}